\documentstyle[12pt,cite]{article}

\newcommand{\nc}{\newcommand}
\nc{\beq}{\begin{equation}} \nc{\eeq}{\end{equation}}
\nc{\beqa}{\begin{eqnarray}} \nc{\eeqa}{\end{eqnarray}}
\nc{\lsim}{\begin{array}{c}\,\sim\vspace{-21pt}\\< \end{array}}
\nc{\gsim}{\begin{array}{c}\sim\vspace{-21pt}\\> \end{array}}
\def\Dslash{\not{\hbox{\kern-3pt $D$}}}

\begin{document}

\title{
{\baselineskip -.2in
\vbox{\small\hskip 4in \hbox{hep-th/0503072}}
\vbox{\small\hskip 4in \hbox{LMU-ASC 06/05}}
\vbox{\small\hskip 4in \hbox{TIFR/TH/05-09}}} 
\vskip .4in
\vbox{
{\bf \LARGE D3 Brane Action and Fermion Zero Modes in Presence of 
Background Flux} 
}
%\narrowtext
\author{Prasanta K. Tripathy${}^1$\thanks{email: prasanta@theorie.physik.uni-muenchen.de}
~and
Sandip P. Trivedi${}^2$\thanks{email: sandip@theory.tifr.res.in} \\
\normalsize{\it
${}^1$ Arnold-Sommerfeld-Center for Theoretical Physics,} \\
\normalsize{\it Department f\"ur Physik, 
Ludwig-Maximilians-Universit\"at M\"unchen,} \\
\normalsize{\it
Theresienstrasse 37, D-80333 M\"unchen, Germany.}  \\
\normalsize{\it ${}^2$ Department of Theoretical Physics,}\\
\normalsize{\it Tata Institute of Fundamental Research,} \\
\normalsize{\it Homi Bhabha Road, Mumbai 400 005, India.}
}}
\maketitle
\begin{abstract}
We derive  the fermion bilinear terms in the world volume action for a $D3$ brane in the 
 presence of background flux. In six-dimensional compactifications non-perturbative 
corrections to the superpotential can arise from an Euclidean D3-brane instanton wrapping 
a divisor  in the internal space. 
The bilinear terms  give rise to fermion masses and 
are important in determining these corrections.  
We find that the three-form flux generically breaks a $U(1)$ subgroup of the structure 
group of the normal bundle of the divisor. In an  example of compactification on
 $T^6/Z_2$,   six of the sixteen zero modes
originally present are lifted by the flux. 

\end{abstract}

\newpage
\section{Introduction}

Flux compactifications have attracted considerable attention recently.
They are of interest from the point of view of string cosmology,
phenomenology, and the general study of string theory vacua with ${\cal
N}\le 1$ supersymmetry. 

Much more needs to be done to understand these compactifications better. 
In particular it should be possible to understand the full
superpotential, including non-perturbative corrections, for these
compactifications in greater depth.  The superpotential has already
proved amazingly useful in the study of supersymmetric string theories
and field theories.  And we can hope that its study for flux
compactifications will prove similarly rewarding. 

An immediate motivation for our work
 is to understand KKLT \cite{kklt} type compactifications better. These
compactifications were first formulated in the context of IIB string
theory of Calabi-Yau orientifolds or related F-theory compactifications.
Here the non-perturbative corrections to the superpotential play a vital
role in stabilising the Kahler moduli \cite{DDF}. 

The study of non-perturbative corrections to the superpotential -in the
closely
 related context of M-theory on a Calabi Yau fourfold - was pioneered by
Witten \cite{witten}. He showed that non-perturbative corrections due to
Euclidean 5-branes wrapping divisors in the four-fold could arise if the
divisor satisfied a particular topological criterion, namely its
arithmetic genus was unity. In Witten's analysis it was
 assumed that a particular $U(1)$ symmetry, which is a  subgroup
of the structure group of the normal bundle to the divisor, was
unbroken. The arithmetic genus is
 an index which counts the number of zero modes after grading by this
symmetry.  More recently, in \cite{GKTT}, a class of non-perturbative
corrections were studied in  a IIB
compactification on $K3 \times T^2/Z_2$, This is related to M theory on
$K3\times K3$. Evidence was found that in the presence of flux the
$U(1)$ symmetry mentioned above is broken. And  it was argued that as a
result non-perturbative corrections could arise even in situations where
the arithmetic genus in not unity.

In this paper we consider a Euclidean D3 brane wrapping a 4-cycle in a
non-trivial background including flux. Using the method of gauge completion we  calculate
all the terms in the action of the D3 brane which are bilinear in
fermions.  These terms explicitly show that the $U(1)$ symmetry of
rotations normal to the 4-cycle is indeed broken in the presence of
flux. As a result the zero mode counting will be altered in general
and modes with the same $U(1)$ charge can pair up and get heavy.  In
a particular example of IIB on an $T^6/Z_2$ orientifold we examine the
resulting fermion zero modes. A non-perturbative correction to the superpotential requires
two zero modes.  Ignoring flux, there are sixteen zero modes.
Including flux, we find for a large  class of divisors that six of the sixteen zero modes are 
lifted. Although this  still leaves  ten zero modes, which is   too many  
 for a correction to be possible, the example
 illustrates the ``efficacy" of flux in lifting  zero-modes.

This paper is only a first step towards a more complete understanding.
One would like to use the results obtained here to understand the number
of zero modes which arise
 more generally. And when the zero mode counting allows for a
correction to the  superpotential,
calculate these corrections. These are
interesting questions which we leave for the future. 

We should also  comment on some of the other relevant literature here. 
We use the method of gauge completion to determine the world volume
theory for the D3 branes. This method is clearly discussed in the paper
by \cite{Crem,deWit}. Our analysis very closely parallels the work by Grana
\cite{Grana}. The only
difference is that we are interested in the more general situation where
the D3 brane is not necessarily transverse to the compactified
directions.  Our results are in agreement after T-duality with those
obtained for the D0 brane by \cite{Millar}. This is a useful  check on
our work. The fermion bilinear terms for a $Dp$ brane in a general background 
 have in fact been obtained earlier in the significant 
papers by Marolf, Martucci and Silva,
\cite{Marolf1,Marolf2}. Our results  can be obtained as a gauge fixed version of their's for the D3-brane case
 and agree. This constitues an important check on our 
results and methods.     
Finally, while we were working on this project, the paper by
Kallosh and Sorokin \cite{Kallosh} appeared which determined the fermion bilinear terms
for an M 5-brane. Using duality  this can be related to the action we calculate here. 
After identifying the relevant gauge
conditions etc we have found substantial agreement \footnote{We thank
R. Kallosh and D. Sorokin for discussions in this regard.}.

This paper is planned as follows. 
The method of gauge completion, which we use to deduce the fermion bilinear terms, is first
briefly  explained in section 2. Following that we illustrate its use for 
some examples and then present the main results determining  the superfields in IIB theory in 
terms of the component supergravity fields upto the required order. 
In section 3 we discuss the resulting D3 brane action.
Our results are checked against those for  a D0 brane using T-duality in section 4, we also comment on the agreement 
with other resulsts in the literature. 
In section 5 we discuss an example of a compactification on a $T^6/Z_2$ orientifold
and calculate the resulting zero modes for a class of divisors. 
Last, but not least, are the six appendices 
which contain some of the important detailed
calculations of the paper.

\section{Gauge Completion}
The approach we will follow for constructing the world volume action of the D3 brane
is straightforward. Given a IIB background in superspace the D3-brane  action 
 can be constructed by appropriately pulling back the background fields
 on to the brane world volume, as is explained in
\cite{ceder1,ceder2,agana1,agana2,Bergsf}. This action has the required supersymmetry and is 
also $\kappa$ symmetric, for on-shell backgrounds. 
We are interested  here 
in  the D3-brane action in terms of the standard component fields of IIB 
supergravity.  So we will first express the superfields of IIB theory in terms of the 
component supergravity  fields by a process called ``gauge completion". 
Once this is done we use the construction mentioned above in terms of the superfields
to obtain the required action.

The method of gauge completion is discussed in \cite{Crem}.
It was applied to the supermembrane in \cite{deWit}. 
Our work will closely parallel the paper of Grana who used an identical strategy.
The only difference is that \cite{Grana} was interested in the case where the D3 brane
is transverse to the compactified directions. We will be interested in obtaining the more 
general answer. Our primary interest is in applying these results to the case of a Euclidean 
D3 brane which wraps a four cycle in the internal directions. In this section and the next two,
where we construct the world volume action and 
compare our results with those obtained in the D0 brane case respectively, we will work in 
Minkowski space. The required transformations to go to Euclidean space will be 
discussed in section \S5 before we apply our results in an explicit example.

The idea behind gauge completion is  to expand the superfields in terms of the fermionic
coordinates, $\theta$, and express each term in the expansion in terms of the 
component fields of supergravity. By the component fields here  we mean the  fields
which appear in the usual  discussion of IIB supergravity, for example, Chapter 13,
 \cite{GSWb} and,
 \cite{Schwarza}. For an on-shell background
these satisfy the equations of motion of the IIB theory.
To lowest order in the $\theta$ expansion of the superfields 
the  component
supergravity fields which appear  are known. To go to higher orders one follows an 
iterative procedure. 
 The idea is that superfields must transform as appropriate
tensors under general bosonic and fermionic coordinate transformations in superspace. 
In particular this included supersymmetry transformations which are translations in the 
fermionic coordinates. Since the supersymmetry transformations for the 
 component supergravity fields are known this allows us to express the higher order 
terms in the $\theta$
expansion in terms of  the lower order ones.  
Obtaining an answer to all orders for a general background in this way 
is computationally quite non-trivial.
Luckily, since we are only interested in terms which are bilinear in the fermions here,
 it will suffice to 
carry out this expansion upto second order in $\theta$ at most.  

In this section we will first 
illustrate this procedure for the dilaton superfield, ${\hat \phi}$
 and the NS-NS two-form superfield, ${\hat B}_{MN}$ and then present the results for
the other superfields towards the end. The calculations are somewhat cumbersome and 
several  details are presented in the Appendix.

Let us begin by explaining our notation. 
We denote superspace coordinates by $Z^M = (x^m,\theta^\mu)$, which stand for the 
bosonic and fermionic components respectively. 
The indices $\{M,N,\cdots\} = \{m,n,\cdots,\mu,\nu,\cdots\}$ denote curved 
(super) coordinates where $(m,n)$ denote Bosonic indices and $(\mu, \nu)$ fermionic indices.
Tangent space indices are given by 
$\{A,B,\cdots\} = \{a,b,\cdots,\alpha,\beta,\cdots\}$, with $(a,b)$ denoting bosonic and 
$(\alpha,\beta)$ fermionic indices.
We will use real $16$ component Majorana-Weyl spinors, our conventions for the Gamma matrices are summarised in Appendix A. 
The spinor indices $\alpha,\beta$ should be viewed as composite indices standing for the 
tensor product of a Majorana-Weyl index and an additional  $SO(2)$ index. 

Our notation for the superfields is as follows. 
A generic superfield  is represented by ${\hat F}_{MN\cdots}$ (with a 
`$~ {\hat{}}~$' over the field). 
The dilaton superfield, whose lowest component is the 
dilaton, $\phi$, is denoted by ${\hat \phi}$, the vierbein superfield by ${\hat{e}_M}^A$
and similarly for ${\hat B}_{MN}, {\hat C}, {\hat C}_{MN}, {\hat C}_{MNPQ}$
which denote the superfields containing the NS-NS two form, and the RR zero, two and four forms respectively. 

Our conventions in superspace are the same as those in \cite{HW}.
Derivatives with respect to $\theta$ are left derivative. Superspace differentials satisfy the property that
$dZ^M\wedge dZ^N=(-1)^{(1+MN)} dZ^N \wedge dZ^M $, where $MN=+1$ when both $M,N$ are fermionic,
and zero otherwise.
A differential two-form for example is given in terms of components by 
${\hat B}=dZ^N dZ^M {\hat B}_{MN}$, and so on. 

Under a  superspace diffeomorphism $Z^M \rightarrow Z^M + \Sigma^M(Z)$, the 
dilaton superfield $\hat\phi$ is a scalar and transforms as 
\begin{eqnarray}
\delta\hat\phi = \Sigma^M\partial_M\hat\phi.
\label{eqphi}
\end{eqnarray}
The fields ${\hat{e}_M}^A$ and ${\hat B}_{MN}$ transform as a vector and a two-index
tensor respectively,
\begin{eqnarray}
&& \delta{\hat{e}_M}^A = \Sigma^P\partial_P{\hat{e}_M}^A
+ \partial_M\Sigma^P {\hat{e}_P}^A \cr
&& \delta{\hat B}_{MN} = \Sigma^P\partial_P{\hat B}_{MN}
+ \left(\partial_M\Sigma^P{\hat B}_{PN} - (-1)^{MN} \partial_N\Sigma^P{\hat B}_{PM}\right)
\label{diffeo}
\end{eqnarray}  ~,
and similarly  for the RR superfields ${\hat C}, {\hat C}_{MN}, {\hat C}_{MNPQ}$. 
We denote the action of a (super) local Lorentz transformation on the vierbein as,
\beq
\label{lorenttrans}
\delta{\hat{e}_M}^A = {\Lambda^A}_B {\hat{e}_M}^B.
\eeq

There are additional  gauge symmetries under which the  the NS-NS two- form and the RR fields
transform, these are superspace generalisations of the familiar gauge symmetries that act on the component supergravity fields.  For example there is a gauge symmetry under which the 
${\hat B}_{MN}$ transforms as,
\beq
\label{local}
 \delta{\hat B}_{MN} = \partial_M\Sigma^{(b)}_N - (-1)^{MN}\partial_N\Sigma^{(b)}_M ~,
\eeq
while the other fields are invariant. Similarly, there are gauge symmetries 
under which  ${\hat C}_{MN}$ and ${\hat C}_{MNPQ}$ transform with gauge transformation
parameters $\Sigma^{(c)}_M$ and $\Sigma^{(c)}_{MNP}$ respectively.

To zeroth order in $\theta$ we have the following identification of the 
superfields in terms of  the component fields.
\begin{eqnarray}
&& {\hat \phi} = \phi \cr
&& {\hat C} = C \cr
&& {\hat{e}_m}^a=e_m^a\cr
&& {\hat{e}_m}^\alpha = {\psi_m}^\alpha \cr
&& {\hat{e}_\mu}^\alpha = {\delta_\mu}^\alpha \cr
&& {\hat B}_{mn} =  B_{mn} \cr
&& {\hat C}_{mn} = C_{mn} \cr
&& {\hat C}_{mnpq} = C_{mnpq} ~,
\label{zeroth}
\end{eqnarray}
and all other fields are zero.

\subsection{The Dilaton Superfield  to $O(\theta^2)$}

We are now ready to illustrate how the procedure of gauge completion works.
We will first examine the dilaton superfield ${\hat \phi}$. 
Consider a super-diffeomorphism which to lowest order in $\theta$ has  components,
\beq
\label{orderotrans}
\Sigma^m=0, \ \ \ \Sigma^\alpha=\epsilon^\alpha.
\eeq
{}From Eq.(\ref{eqphi}) we see that, to $O(\theta^0)$, ${\hat \phi}$ transforms 
under this super-diffeomorphism  as
\begin{eqnarray}
\delta{\hat\phi} % = \Sigma^M\partial_M{\hat\phi}
= \epsilon^\alpha\partial_\alpha{\hat\phi} ~.
\label{dfphi}
\end{eqnarray}
Now since the lowest component of ${\hat \phi}$ is the dilaton, $\phi$, we also know from the 
supersymmetry transformations of IIB supergravity fields (Appendix {\bf A.1}) that 
to this order,
\beq
\label{dphi}
\delta{\hat\phi}=\delta \phi=\bar\epsilon\lambda~.
\eeq
Equating these two expressions  tells us that to $O(\theta^1)$, 
${\hat \phi}$ is given by 
\begin{eqnarray}
\hat\phi = \phi + \bar\theta\lambda ~.
\label{solphi}
\end{eqnarray}

The components of super diffeomorphism we started with, Eq.(\ref{orderotrans}),
are corrected at $O(\theta^1)$. We need to calculate these corrections as the first step 
in obtaining the $O(\theta^2)$ terms in  ${\hat \phi}$.  
This can be done by relating the commutator of two supersymmetry transformations to 
translations. 

Given two supersymmetry transformations with parameters $\epsilon^1, \epsilon^2$
it is straightforward to see that the dilaton transforms under their commutator by a 
translation, 
\begin{eqnarray}
(\delta_1\delta_2 - \delta_2\delta_1)\phi
= \xi^m\partial_m\phi ~,
\label{coma}
\end{eqnarray}
where the translation parameter $\xi^m$ is given by 
\beq
\label{susytrans}
\xi^m = \bar\epsilon_2\Gamma^m\epsilon_1 ~.
\eeq
On the other hand, from Eq.(\ref{eqphi}) we see that the dilaton superfield under the commutator must transform as, 
\begin{eqnarray}
(\delta_1\delta_2 - \delta_2\delta_1)\hat\phi
=\Sigma_2^P\partial_P\Sigma_1^M \partial_M \hat{\phi}
- \Sigma_1^P\partial_P\Sigma_2^M \partial_M \hat{\phi} ~. 
\label{comb}
\end{eqnarray}
Requiring Eq.(\ref{comb}) to agree with Eq.(\ref{susytrans}) upto $O(\theta^0)$ allows us to 
obtain the $O(\theta^1)$ corrections to the super diffeomorphism, Eq.(\ref{orderotrans}). 

We are   interested in this paper in backgrounds  where only bosonic supergravity
 fields acquire 
expectation values and not the  fermionic fields  $\psi_\mu$ and $\lambda$. 
With this in mind, from now on we will set terms depending on fermionic background fields to zero in the 
appropriate  equations. 
To $O(\theta^1)$ one then 
finds that the components of the superdiffeomorphism are given by 
\beq
\label{sdotheta}
\Sigma^m={1\over 2} {\bar \theta} \Gamma^m \epsilon, \ \  \Sigma^\alpha=\epsilon^\alpha.
\eeq
Actually, the 
general solution for $\Sigma^M$ involves certain
undetermined $\theta$ independent tensors. However, as explained elaborately
in \cite{deWit}, by a redefinition of the superspace coordinates we can set
them to zero resulting in Eq.(\ref{sdotheta}). 

The $O(\theta^2)$ terms in the dilaton superfield, ${\hat \phi}$ can now be obtained by 
matching Eq.(\ref{eqphi}) with 
\begin{eqnarray}
\delta\hat\phi = \delta\phi + \bar\theta\delta\lambda ~.
\label{slvphi}
\end{eqnarray}
Using the expression for $\delta\lambda$ as given in the Appendix {\bf A.1}, we find 
\begin{eqnarray}
\hat\phi = \phi + \bar\theta\lambda 
- {1\over 48}  \bar\theta\Gamma^{mnp}\sigma^3\theta H_{mnp} - %\cr &-& 
{1\over 4} e^\phi \bar\theta\Gamma^m(i\sigma^2)\theta F_m
- {1\over 48} e^{\phi} \bar\theta\Gamma^{mnp}\sigma^1\theta F_{mnp} ~.
\label{phislv}
\end{eqnarray}

\subsection{${\hat B}_{MN}$ to $O(\theta^2)$}

We now turn to the NS-NS two-form superfield ${\hat B}_{MN}$. The only new twist here is that 
we will need to include a suitable gauge transformation, Eq.(\ref{local}), with the coordinate transformation, 
Eq.(\ref{sdotheta}),
to determine the $\theta$ expansion in this case. 

To understand this let us first calculate the commutator of two supersymmetry transformations, with parameters,
$\epsilon^1, \epsilon^2$ on the component supergravity field $B_{mn}$.  Using the susy transformation rules given in the Appendix {\bf A.1},
\begin{eqnarray}
\delta_1\delta_2 B_{mn} = \bar\epsilon_2\sigma^3\left(
\Gamma_m\delta_1\psi_n - \Gamma_n\delta_1\psi_m\right) ~,
\end{eqnarray}
we find that 
\begin{eqnarray}
(\delta_1\delta_2 - \delta_2\delta_1) B_{mn}
&=& - \left(\partial_m(\bar\epsilon_2\sigma^3\Gamma_n\epsilon_1)
- \partial_n(\bar\epsilon_2\sigma^3\Gamma_m\epsilon_1) \right) 
+ \bar\epsilon_2\Gamma^p\epsilon_1 H_{mnp} \cr
&=& \xi^p\partial_p B_{mn} + \partial_m\xi^p B_{pn} - \partial_n\xi^p B_{pm}
+ \partial_m\xi^{12(b)}_n - \partial_n\xi^{12(b)}_m ~.
\label{dlta}
\end{eqnarray}
The second line on the r.h.s. is the transformation of $B_{mn}$ under a combined  translation by $\xi^m$ and a 
gauge transformation with parameter $\xi^{12(b)}_n$.
One finds that this equation can be met if $\xi^n$ is given by Eq.(\ref{susytrans}), and the gauge 
transformation parameter is, 
\begin{eqnarray}
\xi^{12(b)}_m = \xi^n B_{mn} - \bar\epsilon_2\sigma^3\Gamma_m\epsilon_1 ~.
\label{xib}
\end{eqnarray}

In terms of superfields this tells us that the super-diffeomorphism, Eq.(\ref{sdotheta}), should 
be accompanied by a gauge transformation.
We denote the gauge transformation parameter in superspace by $\Sigma^{(b)}$ below. The combined transformation can 
then be written as,
\beq
\label{comtransb}
\delta{\hat B}_{MN}=\Sigma^P\partial_P {\hat B}_{MN}+\partial_M\Sigma^PB_{PN}-(-1)^{MN}\partial_N\Sigma^PB_{PM}
+\partial_M \Sigma^{(b)}_N-(-1)^{MN}\partial_N\Sigma^{(b)}_M~.
\eeq
 
The commutator of two  transformations in superspace can now  be  calculated. 
We get that 
\begin{eqnarray}
(\delta_1\delta_2 - \delta_2\delta_1){\hat B}_{MN}
= \partial_M\Sigma_N^{12(b)} -(-1)^{MN} \partial_N\Sigma_M^{12(b)} + \cdots  ~.
\label{onetwo}
\end{eqnarray}
The ellipses on the rhs denote extra terms which arise due to a coordinate transformation with parameters, 
Eq.(\ref{sdotheta}).
$\Sigma^{12(b)}$ above denotes a gauge transformation, it's components turn out to be, 
\beq
\label{xiontw}
\Sigma_M^{12(b)} = \left(\Sigma_2^P\partial_P\Sigma^{1(b)}_M  
 + \partial_M\Sigma_2^P \Sigma^{1(b)}_P\right) 
- \left(1\leftrightarrow 2\right) ~.
\eeq

To leading order in $\theta$, $B_{m\mu}$ and $B_{\mu\nu}$ both vanish and the only non-zero component 
of ${\hat B}_{MN}$ is $B_{mn}$.
Comparing Eq.(\ref{xiontw}) and Eq.(\ref{xib}) and using Eq.(\ref{sdotheta}) for the components of $\Sigma^P$ we 
then find that  upto $O(\theta)$,
\begin{eqnarray}
\Sigma_m^{(b)} &=&  {1\over 2} \bar\theta\left(
\Gamma^n B_{mn} - \sigma^3\Gamma_m\right)\epsilon ~.
\label{sigmamb}
\end{eqnarray}
And $\Sigma^{(b)}_\mu=0$.  

We are now ready to evaluate $\hat{B}_{MN}$ to higher orders in $\theta$. From the susy transformation,
Appendix {\bf A.1}, 
for the supergravity field $B_{mn}$ it follows that upto $O(\theta^1)$
\begin{eqnarray}
{\hat B}_{mn} = B_{mn} + \bar\theta\sigma^3\Gamma_m\psi_n - \bar\theta\sigma^3\Gamma_n\psi_m
~. 
\label{bmnf}
\end{eqnarray}
To evaluate ${\hat B}_{m\mu}$, note that
\begin{eqnarray}
\delta{\hat B}_{m\mu} = \epsilon^\alpha\partial_\alpha{\hat B}_{m\mu}
+ {1\over 2} \left(\bar\epsilon\sigma^3\Gamma_m\right)_\mu
\label{bmmu}
\end{eqnarray}
Since $B_{m\mu}$ vanishes at zeroth order in $\theta$, the 
above variation should be zero, which gives
\begin{eqnarray}
{\hat B}_{m\mu} = - {1\over 2} \left(\bar\theta\sigma^3\Gamma_m\right)_\mu ~.
\label{bmum}
\end{eqnarray}
Similarly one can show that ${\hat B}_{\mu\nu}$ must vanish upto $O(\theta^1)$. 

To find the second order results for ${\hat B}_{mn}$, we consider
the variation of ${\hat B}_{mn}$, Eq.(\ref{comtransb}), upto  to first order in $\theta$. 
Using the results for the superdiffeomorphism, Eq.(\ref{sdotheta}), and gauge transformation, Eq.(\ref{sigmamb}),
this is given by  
\begin{eqnarray}
\delta{\hat B}_{mn} &=& \epsilon^\alpha\partial_\alpha{\hat B}_{mn}  
+ {1\over 2} \bar\theta\Gamma^p\epsilon H_{mnp}  \cr
&+& \bar\theta\sigma^3\left(\Gamma_m\partial_n\epsilon
- \Gamma_n\partial_m\epsilon\right)
- {1\over 2} \bar\theta\sigma^3\Gamma_a\epsilon
\left(\partial_m{e_n}^a - \partial_n{e_m}^a \right) ~.
\label{bmndlt}
\end{eqnarray}
On the other hand this has to be equated with the variation of Eq.(\ref{bmnf}) leading to  
\begin{eqnarray}
\delta{\hat B}_{mn} &=& \delta{B}_{mn} + \bar\theta\sigma^3
\left(\Gamma_m\delta\psi_n - \Gamma_n\delta\psi_m \right) \cr
&=& \bar\epsilon\sigma^3\left(\Gamma_m\psi_n - \Gamma_n\psi_m\right)
+ \bar\theta\sigma^3
\left(\Gamma_m\partial_n\epsilon - \Gamma_n\partial_m\epsilon\right) - %\cr &-& 
{1\over 2}\bar\theta\sigma^3\Gamma_a\epsilon
\left(\partial_m{e_n}^a - \partial_n{e_m}^a\right) \cr &+&
 {1\over 4}\bar\theta\sigma^3\left({\omega_m}^{ab}\Gamma_{nab}
- {\omega_n}^{ab}\Gamma_{mab}\right)\epsilon - %\cr &-& 
{1\over 4} e^\phi \bar\theta\sigma^1\Gamma_{mnp}\epsilon F^p
+ {1\over 2} \bar\theta\Gamma^p\epsilon H_{mnp} \cr &-&
 {1\over 8} \bar\theta\left(
{\Gamma_m}^{pq} H_{npq} - {\Gamma_m}^{pq} H_{mpq} \right)\epsilon - %\cr &-& 
{1\over 24} e^\phi \bar\theta(i\sigma^2)\left(
{\Gamma_{mn}}^{pqr} F'_{pqr} + 6 \Gamma^p F'_{mnp}\right)\epsilon \cr &-& 
{1\over 8\cdot 5!} e^\phi \bar\theta\sigma^1
\left({\Gamma_{mn}}^{pqrst} F'_{pqrst} + 20 \Gamma^{pqr} F'_{mnpqr}\right)
\epsilon ~,
\label{dltbmn}
\end{eqnarray}
where on the rhs we have used the susy transformations for $B_{mn}$ and the gravitino from Appendix {\bf A.1}. 
Eq.(\ref{bmndlt}), (\ref{dltbmn}) finally give us the expansion to second order in $\theta$,
\begin{eqnarray}
{\hat B}_{mn} &=& B_{mn} 
+ \bar\theta\sigma^3\left(\Gamma_m\psi_n - \Gamma_n\psi_m \right) \cr
&+& {1\over 8}\bar\theta\sigma^3\left({\omega_m}^{ab}\Gamma_{nab}
- {\omega_n}^{ab}\Gamma_{mab}\right)\theta 
- {1\over 16} \bar\theta\left(
{\Gamma_m}^{pq} H_{npq} - {\Gamma_m}^{pq} H_{mpq} \right)\theta \cr
&-& {1\over 8} e^\phi \bar\theta\sigma^1\Gamma_{mnp}\theta F^p
- {1\over 48} e^\phi \bar\theta(i\sigma^2)\left(
{\Gamma_{mn}}^{pqr} F'_{pqr} + 6 \Gamma^p F'_{mnp}\right)\theta \cr
&-& {1\over 16\cdot 5!} e^\phi \bar\theta\sigma^1
\left({\Gamma_{mn}}^{pqrst} F'_{pqrst} + 20 \Gamma^{pqr} F'_{mnpqr}\right)
\theta ~.
\label{bmn}
\end{eqnarray}

As was mentioned in the discussion of the previous subsection we are interested in backgrounds for which the 
 fermions $\psi_m$ and $\lambda$ are zero. Also, when we construct the world volume action it will be convenient to
 work in static  gauge and fix the $\kappa$-symmetry  
by choosing the space time spinors $\theta_1,\theta_2$ to be 
\begin{eqnarray}
\theta_1 = \Theta ~,~~ \theta_2 = 0 ~.
\label{theta}
\end{eqnarray}
With this choice the expression for ${\hat B}_{mn}$ becomes
\begin{eqnarray}
{\hat B}_{mn}= B_{mn}
+ {1\over 8}\bar\Theta\left({\omega_m}^{ab}\Gamma_{nab}
- {\omega_n}^{ab}\Gamma_{mab}\right)\Theta
- {1\over 16} \bar\Theta\left(
{\Gamma_m}^{pq} H_{npq} - {\Gamma_m}^{pq} H_{mpq} \right)\Theta ~.
\label{bhat}
\end{eqnarray}

It will be enough for our purposes of determining the fermion bilinear terms below 
to determine the other components $B_{m\mu}, B_{\mu\nu}$, to $O(\theta^1)$ which was already done above.

\subsection{Final Results}

One can follow through similar steps to obtain the expansions for 
other superfield. We have summarised the results below, detail calculations
are performed in the Appendix.

As was mentioned above, we have set the fermionic backgrounds to zero.
Also we work with the choice of spinors in Eq.(\ref{theta}). 
The  components of the superfields to required order in the $\theta$ expansion are then
 given by: 
\begin{eqnarray}
\hat\phi &=& \phi - {1\over 48}  \bar\Theta\Gamma^{mnp}\Theta H_{mnp} \cr
\hat{C} &=&  C - {1\over 48} \bar\Theta \Gamma^{mnp} \Theta F'_{mnp} \cr
{{\hat e}_\mu}^a &=& - {1\over 2} \left(\bar\theta\Gamma^a\right)_\mu\cr
{{\hat e}_m}^a &=& {e_m}^a 
- {1\over 8} {\omega_m}^{cd}\bar\Theta{\Gamma^a}_{cd} \Theta
- {1\over 16}  H_{mpq} \bar\Theta\Gamma^{apq}\Theta \cr
\hat{B}_{m\mu} &=& - {1\over 2} \left(\bar\theta\sigma^3\Gamma_m\right)_\mu \cr
\hat{B}_{mn} &=& B_{mn} 
- {1\over 8}  \bar\Theta\left({\Gamma_m}^{ab}\omega_{nab}
- {\Gamma_n}^{ab}\omega_{mab}\right)\Theta -  %\cr &-& 
{1\over 16} \bar\Theta\left({\Gamma_m}^{pq} H_{npq} 
                             - {\Gamma_n}^{pq} H_{mpq}\right)\Theta \cr
\hat{C}_{m\mu} 
&=& {1\over 2} e^{-\phi} \left(\bar\theta\sigma^1\Gamma_m\right)_\mu
  - {1\over 2} C \left(\bar\theta\sigma^3\Gamma_m\right)_\mu \cr
\hat{C}_{mn} &=& C_{mn}
- {1\over 8}  C \bar\Theta\left({\Gamma_m}^{ab}\omega_{nab}
- {\Gamma_n}^{ab}\omega_{mab}\right)\Theta 
+ {1\over 8}  \bar\Theta\Gamma_{mnp}\Theta F^p \cr &-&
 {1\over 16} C \bar\Theta\left({\Gamma_m}^{pq} H_{npq} 
                               - {\Gamma_n}^{pq} H_{mpq}\right)\Theta 
- {1\over 16} \bar\Theta\left({\Gamma_m}^{pq} F'_{npq} 
                               - {\Gamma_n}^{pq} F'_{mpq}\right)\Theta \cr
&-& {1\over 16\cdot 5!}  \bar\Theta\left(
{\Gamma_{mn}}^{pqrst} F'_{pqrst} + 20 \Gamma^{pqr} F'_{mnpqr}\right)\Theta \cr
\hat{C}_{\mu mnp} &=& 
- {1\over 2} e^{-\phi} \left(\bar\theta(i\sigma^2)\Gamma_{mnp}\right)_\mu
+ {3\over 2} \left(\bar\theta\sigma^3 C_{[mn} \Gamma_{p]}\right)_\mu \cr
\hat{C}_{mnpq} &=& C_{mnpq} 
- {3\over 2}  \bar\Theta C_{[mn}{\Gamma_p}^{ab}\omega_{q]ab}\Theta
- {3\over 4} \bar\Theta C_{[mn} {\Gamma_p}^{st} H_{q]st}\Theta 
+ \bar\Theta\left( {1\over 48}  {\Gamma_{mnpq}}^{stu}F'_{stu}
\right. \cr &+& 
\left. {1\over 2} \Gamma_{[mnp} F_{q]}%\Theta
+ {3\over 4} {\Gamma_{[mn}}^s F'_{pq]s} - %\cr &-& 
{1\over 96} {\Gamma_{[mnp}}^{stuv} F'_{q]stuv} 
%\cr &-& 
-{1\over 8} {\Gamma_{[m}}^{st} F'_{npq]st}\right)\Theta ~.
\label{summary}
\end{eqnarray}
Here, $H_3=dB$.  And $F'_{mnp}$, $F'_{mnpqrs},$ refer  to the components of the three form, 
$dC_2-C_0H_3$, and the five form, $dC_4-C_2\wedge H_3$, respectively. 
Eq.(\ref{summary})  is one of the main results of our paper. 

\section{World Volume Action}
\subsection{The Action}
The action for the $D3$ brane is given by 
\cite{ceder1,ceder2,agana1,agana2,Bergsf}
\begin{eqnarray}
S = - \mu_3\int d^4\zeta e^{-\hat\phi} \sqrt{-{\rm det}\left(
{\hat g}_{\bf \tilde i\tilde j} + F_{\tilde i\tilde j} - {\hat B}_{\bf \tilde i\tilde j}\right)}
+ \mu_3 \int e^{F - {\hat B}}\wedge {\bf \hat C} ~.
\label{dbics}
\end{eqnarray}
It is obtained by pulling back the superfields from spacetime to the $D3$ brane world volume.
For on-shell background fields the action is $\kappa$-symmetric.
In Eq.(\ref{dbics}) $\zeta^{\tilde i}, \tilde i=0, \cdots 3 $ are the world volume coordinate. 
We also  denote ${\bf\hat C} = \oplus_n {\hat C}_n$. 

It will be useful in the discussion below
to distinguish between the pullback of the superfield and pullback of the component bosonic
supergravity fields. 
The pullback of a  superfield is  by definition obtained by pulling back the superspace tensor
onto the worldvolume. For example, the pullback of ${\hat B}_{MN}$ is,
\begin{eqnarray}
{\hat B}_{\bf \tilde i\tilde j} 
= \partial_{\tilde i}Z^M\partial_{\tilde j}Z^N {\hat B}_{MN} ~,
\label{psb}
\end{eqnarray}
where $Z^M=(x^m, \theta^\mu)$ are the spacetime superspace coordinates. This is what appears
in Eq.(\ref{dbics}). 
In contrast we define the pullback of the component supergravity field from the 
ordinary (Bosonic)
target space to the worldvolume. So,
\begin{eqnarray}
B_{\tilde i\tilde j} = \partial_{\tilde i}x^m\partial_{\tilde j}x^n B_{mn} ~.
\end{eqnarray}
To distinguish between the two we will use boldface indices in the case of the superfield,
as in Eq.(\ref{dbics}), (\ref{psb}) above. 

It will also be useful to distinguish between the lowest order term and the higher order
contributions in the $\theta$ expansion for any superfield. the latter will be denoted
by an additional prime.  
For example, we can write for the dilaton superfield,
\beq
\label{notphi}
{\hat \phi}=\phi + \phi'
\eeq
where from Eq.(\ref{summary}), 
$\phi'=- {1\over 48}  \bar\Theta\Gamma^{mnp}\Theta H_{mnp}$ .

Using the expressions for the super vierbeins from Eq.(\ref{summary}), it then follows that 
 the metric 
${\hat g}_{\bf \tilde i\tilde j} = {\hat e_{\bf \tilde i}}^a {\hat e_{\bf \tilde j}}^b \eta_{ab}$ to
second order in $\Theta$ is,
\begin{equation}
\hat{g}_{\bf \tilde i\tilde j}=g_{\tilde i\tilde j}
+\left(\partial_{\tilde i}x^m\partial_{\tilde j}x^n
+\partial_{\tilde j}x^m\partial_{\tilde i}x^n\right)e_n^b{e'}^a_m\eta_{ab}
+{1\over 2}\bar\Theta\Gamma^a\left(D_{\tilde i}\Theta\partial_{\tilde j}x^m
+D_{\tilde j}\Theta\partial_{\tilde i}x^m\right)e_m^b\eta_{ab}~.
\label{metricpb}
\end{equation}
A similar straightforward analysis shows that the pull back of the 
NS and RR superfields become
\begin{eqnarray}
\hat{B}_{\bf \tilde i\tilde j}  &=&  B_{\tilde i\tilde j} 
+ \partial_{\tilde i} x^m \partial_{\tilde j} x^n B'_{mn}
+ {1\over 2}  \left(\bar\Theta\Gamma_{\tilde i} D_{\tilde j}\Theta
- \bar\Theta\Gamma_{\tilde j} D_{\tilde i}\Theta\right) \cr
{\hat C}_{\bf \tilde i\tilde j} &=&  C_{\tilde i\tilde j} 
+ \partial_{\tilde i} x^m \partial_{\tilde j} x^n C'_{mn}
+ {1\over 2} C \left(\bar\Theta\Gamma_{\tilde i}\partial_{\tilde j}\Theta
- \bar\Theta\Gamma_{\tilde j}\partial_{\tilde i}\Theta\right) \cr
{\hat C}_{\bf \tilde i\tilde j\tilde k\tilde l} 
&=& C_{\tilde i\tilde j\tilde k\tilde l}
+ \partial_{\tilde i} x^m \partial_{\tilde j} x^n 
\partial_{\tilde k} x^p \partial_{\tilde l} x^q C'_{mnpq}
+ 4! \partial_{[\tilde i}\Theta^\mu \partial_{\tilde j} x^n 
\partial_{\tilde k} x^p \partial_{\tilde l]} x^q
         {\hat C}_{\mu npq} ~.
\label{pullbacks}
\end{eqnarray}

Using these expressions we can compute the world volume action. The DBI
action becomes
\begin{eqnarray}
S_{DBI} = - \mu_3\int d^4\zeta e^{-\phi} \sqrt{{\rm det} g}
\left\{\left(1 + {1\over 4} (F - B)^2\right)
\left(1+ {1\over 48} \bar\Theta\Gamma^{mnp}\Theta H_{mnp}\right)\right. \cr
\left. + {1\over 2} \left({\delta_{\tilde i}}^{\tilde k} 
+ {(F-B)_{\tilde i}}^{\tilde k}\right)
\left(\bar\Theta\Gamma_{\tilde k}D^{\tilde i}\Theta
- {1\over 8}\bar\Theta\Gamma_{\tilde kpq}\Theta H^{\tilde ipq}\right) +\cdots \right\} ~.
\label{sdbi}
\end{eqnarray}
Here we have followed a slightly condensed notation.
In our notation above, $\tilde i,\tilde j, \tilde k$ denote world volume indices, whereas $m,n,p$ denote 
spacetime (bosonic) indices. Now, 
$\Gamma_{\tilde k} \equiv\Gamma_m \partial_{\tilde k}x^m$,  
$\partial^{\tilde i}\Theta \equiv g^{\tilde i\tilde j}\partial_{\tilde j}\Theta$, etc.
The ellipses on the right hand side above indicate additional terms that can be obtained by
 expanding the 
square root in Eq.(\ref{dbics}) to higher order. In addition  of course extra terms  
would arise if we carried out  the $\theta$ expansion of the superfields to higher order as
 well.

Similarly the  
 Wess-Zumino action is 
\begin{eqnarray}
S_{WZ} &=&  \mu_3 \int e^{(F - B)} \wedge {\bf C}
- {1\over 96} \mu_3\int (F-B)\wedge(F-B)
\bar\Theta\Gamma^{mnp}\Theta F'_{mnp} \cr
 &+& {1\over 32}\mu_3\int d^4\zeta \sqrt{{\rm det} g} 
\epsilon^{\tilde i\tilde j\tilde k\tilde l} (F-B)_{\tilde i\tilde j} \bar\Theta
\left\{\Gamma_{\tilde k\tilde lp} F^p - {\Gamma_{\tilde k}}^{pq} F'_{\tilde lpq}
%\right. \cr &-& \left. 
- {1\over 2\cdot 5!} \left(
{\Gamma_{\tilde k\tilde l}}^{pqrst} F'_{pqrst} 
\right.\right. \cr &+& \left.\left. 
 20 \Gamma^{pqr} F'_{\tilde k\tilde lpqr}\right)
\right\} \Theta +
{1\over 16} \mu_3\int d^4\zeta \sqrt{{\rm det} g}  \epsilon^{\tilde i\tilde j\tilde k\tilde l}  \left(
{1\over 72} \bar\Theta{\Gamma_{\tilde i\tilde j\tilde k\tilde l}}^{pqr}\Theta F'_{pqr}
+ {1\over 3} \bar\Theta\Gamma_{\tilde i\tilde j\tilde k}\Theta F_{\tilde l}
\right. \cr & + & \left.
 {1\over 2} \bar\Theta{\Gamma_{\tilde i\tilde j}}^p\Theta F'_{\tilde k\tilde lp}
- {1\over 3!} \bar\Theta{\Gamma_{\tilde i}}^{pq}\Theta F'_{\tilde j\tilde k\tilde lpq}
\right) ~.
\label{wzcs}
\end{eqnarray}
Equations, (\ref{sdbi}) and (\ref{wzcs}), are important for the the following discussion. 
We will see in the next section that the action above agrees with other established results in the literature.

\subsection{ Some Comments}

Two comments are now in order.

One of the main motivations for this project is to understanding non-perturbative corrections
to the superpotential which can arise in flux compactifications. In this context we are 
interested  in IIB string theory compactified down to $R^{3,1}$ (actually 
for the non-perturbative 
corrections we are interested in the Euclidean situation $R^4$ as discussed in the next 
section). One class of non-perturbative effects, which is our main focus here, arise due to 
to Euclidean D3 branes that wrap a  holomorphic 4-cycle, i.e. a divisor,
 in the internal 6-dimensional space. 

Under a duality map to M-theory this   lifts to a Euclidean 5-brane instanton wrapping a 
divisor of the Calabi-Yau four-fold. The resulting superpotential was discussed in the 
seminal paper of Witten \cite{witten}.
An $U(1)$ symmetry played an important role in this analysis. 
This symmetry is a subgroup of the structure group of the normal bundle and 
corresponds to rotations in the plane of the 
two compact directions orthogonal to the divisor. An index was formulated 
 by counting the fermionic zero modes after grading them by their
 charge under this symmetry. This index turned out to be proportional to the arithmetic
genus of the divisor and it was  argued that a correction could only arise if the
arithmetic genus was unity.

In the IIB description we are using here the $U(1)$ the divisor is $2$ complex dimensional
and the compactified space is 6-dimensional. This means, roughly speaking, that two compact directions are normal to the divisor and the $U(1)$ symmetry is rotations in the plane
formed by these two directions. 
We will now see that the presence of three-form flux can lead to
 this $U(1)$ symmetry being broken in the D3-brane world volume theory. As a result, zero
 modes with the {\it same}  $U(1)$ charge can   pair up and get heavy.
In this way, a correction to the superpotential  can  arise even though the index 
condition mentioned above is not met.

The essential point is simply that if the three form flux has two legs along the 4-cycle and 
one perpendicular to it then it will break the $U(1)$ symmetry mentioned above. 
Since the fluxes enter in various bilinear fermion couplings in 
Eqs.(\ref{sdbi}) and (\ref{wzcs}), the mass
terms for the fermions will in general violate this symmetry. To illustrate this concretely
let us consider the situation where $F-B$ in the world-volume theory vanishes.
Then the fermion three-form flux dependent 
mass  terms for a D3 brane wrapping a 4-cycle take the form, 
\begin{eqnarray}
\label{simpleact}
S_{Mass}=-\mu_3\int d^4\xi \sqrt{{\rm det}g} 
\bar\Theta\left\{e^{-\phi}{1\over 48}\Gamma^{mnp} H_{mnp}  
- {1\over 16} e^{-\phi}\Gamma_{\tilde ipq} H^{\tilde ipq} 
-{1\over 32}\epsilon^{\tilde i\tilde j\tilde k\tilde l} \Gamma_{\tilde i\tilde j}^pF'_{\tilde k\tilde lp} \right\}\Theta 
\end{eqnarray} 
(we have used the fact that the flux preserves Poincare invariance in $R^{3,1}$ to set
some terms to zero). We remind the reader that in our notation, indices, $\tilde i,\tilde j,\tilde k,\tilde l$ are 
along the worldvolume, and $m,n,p$ take $0, \cdots 9$ values in spacetime. 
Now, in general, it is easy to see that if $H,F'$ have two legs along the brane and one along 
the normal then each of the terms appearing above 
breaks the $U(1)$ symmetry. Also, the sum of these terms does not vanish 
for on-shell backgrounds, even those which meet the conditions of supersymmetry. 
Thus, as was mentioned above the mass terms will in general break the $U(1)$ symmetry allowing
in particular two fermions with same sign charges to pair up and get heavy.  
 
Second, let us now consider the special case of   a $D3$-brane which is along $R^{3,1}$ and 
transverse to the internal directions. We also take the background fields to preserve the 
Poincare symmetry of $R^{3,1}$. In addition, we take the space time metric 
to be of the form $g_{10} = e^{2A(y^m)}\eta_{4}\otimes g_6^{tr}$.
The DBI term is then given by,
\begin{eqnarray}
S_{DBI} &=& - \mu_3\int d^4\zeta e^{-\phi} \sqrt{{\rm det} g}
\left\{\left(1 + {1\over 4} (F-B)^2\right)
\left(1+ {1\over 48} \bar\Theta\Gamma^{mnp}\Theta H_{mnp}\right)\right. \cr
&& ~~~~~~~~~~~~~~~~~~~~~~~~~~~ 
\left. + {1\over 2} \left({\delta_{\tilde i}}^{\tilde k} 
+ {(F-B)_{\tilde i}}^{\tilde k}\right)
\bar\Theta\Gamma_{\tilde k}\partial^{\tilde i}\Theta
 +\cdots \right\} ~.
\label{dibor}
\end{eqnarray}
The spin connection dependent term vanishes in the above equation for the 
a general warped metric. The Wess-Zumino term is given by
\begin{eqnarray}
\label{newwz}
S_{WZ} &=&  \mu_3 \int  C_4
- {1\over 96} \mu_3\int (F-B)\wedge(F-B)
\bar\Theta\Gamma^{mnp}\Theta F'_{mnp} \cr &+&
  {1\over 32}\mu_3\int d^4\zeta \sqrt{{\rm det} g} 
\epsilon^{\tilde i\tilde j\tilde k\tilde l}  (F-B)_{\tilde i\tilde j} 
\bar\Theta\left\{
%\right. \cr && \left.
 \Gamma_{\tilde k\tilde lp} F^p - 
  {1\over 2\cdot 5!} \left({\Gamma_{\tilde k\tilde l}}^{pqrst} F'_{pqrst}
 \right.\right. \cr &+& \left.\left. 
 20 \Gamma^{pqr}F'_{klpqr}\right)\right\}\Theta +
{1\over 48\cdot 4!} \mu_3\int d^4\zeta \sqrt{{\rm det} g} \epsilon^{\tilde i\tilde j\tilde k\tilde l} 
\bar\Theta{\Gamma_{\tilde i\tilde j\tilde k\tilde l}}^{pqr}\Theta F'_{pqr} ~.
\end{eqnarray}
The full action is the sum of these two terms. 
This result is of interest from the point of view of calculating the soft terms that can arise
after turning on fluxes  \cite{soft1,soft2,soft3,soft4,soft5,soft6,soft7,soft8}.
It   agrees (upto some minor discrepancy in the numerical factors)
with Ref.~\cite{Grana}. 

Ignoring terms dependent on $(F-B)$, the $O(\Theta^2)$ part of the action
becomes
\begin{equation}
S(\Theta^2) = {\mu_3\over 48} \int d^4\zeta e^{-\phi} \sqrt{{\rm det} g}
\bar\Theta\Gamma^{mnp}\Theta {\rm Re}(*G - i G)_{mnp} ~,
\end{equation}
where $G \equiv F' - i e^{-\phi} H$. We see that for imaginary self dual
flux, the above term vanishes. This is to be expected from the analysis of \cite{GKP}.

\section{T-duality And Comparison With Other Results}

As a simple check of our results, we can take the type IIA action
for D0 brane and perform three T-dualities to obtain the action for D3 brane. 
The D0 brane action to order $\Theta^2$, in the Einstein frame,
 is given by \cite{Millar}
\begin{eqnarray}
S &=& - \mu_0 \int d\tau e^{-{3\over 4}\phi} 
\left(1 - {3\over 4} \Phi|_{\Theta^2} + \cdots \right)
\sqrt{ - \left(g_{mn} + 2 e_{m a} E_n^a|_{\Theta^2} + \cdots\right)
\dot{x}^m\dot{x}^n} \cr
&+& \mu_0 \int d\tau \left( C_m + B_m|_{\Theta^2} + \cdots \right) \dot{x}^m
\label{dzero}
\end{eqnarray}
where the dots indicate terms of higher order in $\Theta$.
The order $\Theta^2$ part of the IIA superfields are given by
\begin{eqnarray}
&& \Phi|_{\Theta^2} = {i\over 48} e^{-{1\over 2}\phi} 
\bar\Theta\Gamma^{mnp}\Theta G_{mnp} \cr
&& B_{m}|_{\Theta^2} = - {i\over 16} 
\bar\Theta{\gamma_m}^{np}\Theta F_{np}
- {i\over 48} e^{-{1\over 2}\phi}
\bar\Theta\gamma^{npq} F'_{mnpq} \cr
&&  
E_m^a|_{\Theta^2} = {i\over 8} \bar\Theta\gamma^{abc}\Theta\omega_{mbc}\cr
&& ~~~~~~~~ + {i\over 64} e^{-{1\over 2}\phi}\left(
\bar\Theta{\gamma_m}^{np}\Theta {H^r}_{np}
+ 3 \bar\Theta\gamma^{anp}\Theta H_{mnp}
- {1\over 3} e_m^a \bar\Theta\gamma^{npq}\Theta H_{npq}
\right)
\label{zerod}
\end{eqnarray}
Using the above formulae, we write the action in terms component fields. 
Also, before performing T-duality, we make the following 
field redefinitions \cite{Grana} to change the action in to sting frame. 
\begin{eqnarray}
g_{mn(E)} = e^{-{1\over 2}\phi} g_{mn(S)} ~, ~~
\Gamma^m_{(E)} = e^{{1\over 4}\phi} \Gamma^m_{(S)} ~, ~~
\Theta_{(E)} = e^{-{1\over 8}\phi} \Theta_{(S)} ~.
\end{eqnarray}
With this, the DBI action becomes
%\begin{eqnarray}
%S_{DBI}=-\mu_0\int d\tau e^{-\phi}
%\sqrt{-g_{00}}\left(1+{i\over 8}
%\left\{\bar\Theta\gamma^{0ab}\Theta\omega_{0ab}
%+{1\over 2} \bar\Theta\gamma^{0np}
%\Theta H_{0np}
%\right.\right. \cr  \left.\left. 
%-{1\over 6} \bar\Theta\gamma^{mnp}\Theta H_{mnp}
%\right\}+\cdots\right)
%\label{dbizero}
%\end{eqnarray}
\begin{equation}
S_{DBI}=-\mu_0\int d\tau e^{-\phi}
\sqrt{-g_{00}}\left(1+{i\over 8}\bar\Theta\left\{\gamma^{0ab}\omega_{0ab}
+{1\over 2} \gamma^{0np} H_{0np}
-{1\over 6} \gamma^{mnp} H_{mnp}\right\}\Theta+\cdots\right)
\label{dbizero}
\end{equation}
and the Wess-Zumino part 
\begin{eqnarray}
{S_{WZ}} = \mu_0 \int d\tau \left\{ C_0 
-{i\over 16} \left(\bar\Theta{\gamma_0}^{mn}\Theta F_{mn}
+ {1\over 3} \bar\Theta{\gamma}^{mnp}\Theta F'_{0mnp}
\right) + \cdots \right\}
\label{wzzero}
\end{eqnarray}

Now we perform three T-dualities along $\{x,y,z\}$. Let us denote
these directions by $\acute{m},\acute{n},\cdots$ 
and the remaining directions by 
$\check{p},\check{q}, \cdots$. For simplicity, we consider the following special case. 
We assume 
$g_{\acute m\check p} = B_{\acute n\check q} =B_{\acute n\acute m} = 0$ and  
we take the metric along the directions
$x,y,z$ to be diagonal.  Also we set the spin 
connection to zero. Using the T-duality rules as given in the Appendix 
{\bf A.5},
it is then  straightforward to see that the quadratic part of the action 
(\ref{dbizero}) is identical to our result
\begin{eqnarray}
S_{DBI}(\Theta^2) = - \mu_3\int d^4\zeta e^{-\phi} \sqrt{{\rm det}~g}
\left({1\over 48} \bar\Theta\Gamma^{mnp}\Theta H_{mnp}
- {1\over 16} \bar\Theta\Gamma_{\tilde ipq}\Theta H^{\tilde ipq}
\right)
\end{eqnarray}
We can now turn to the quadratic part of the Wess-Zumino action. After 
performing the duality, we find
\begin{eqnarray}
&-&{i\over 16} \left(\bar\Theta{\gamma_0}^{mn}\Theta F_{mn}
+ {1\over 3} \bar\Theta{\gamma}^{mnp}\Theta F'_{0mnp}
\right) \cr
&=&  \bar\Theta \left\{
{i\over 16} {\gamma_0}^{\check p\check q} F_{xyz\check p\check q}
+ {i\over 8} \left(
{\gamma_{0x}}^{\check p} F_{yz\check p}
+ {\gamma_{0y}}^{\check p} F_{zx\check p}
+ {\gamma_{0z}}^{\check p} F_{xy\check p}
\right)
- {i\over 48} \gamma^{\check p\check q\check r} 
F'_{0xyz\check p\check q\check r}
\right. \cr &+& \left.
{i\over 8} \left( \gamma_{xyz} F_0
- \gamma_{0xy} F_{z} - \gamma_{0yz} F_x - \gamma_{0zx} F_y \right)
- {i\over 16} \left(
 {\gamma_x}^{\check p\check q} F'_{0yz\check p\check q}
+ {\gamma_y}^{\check p\check q} F'_{0zx\check p\check q}
\right.\right.\cr &+& \left.\left.
{\gamma_z}^{\check p\check q} F'_{0xy\check p\check q}
\right) 
- {i\over 8} \left(
{\gamma_{xy}}^{\check p} F'_{z0\check p}
+ {\gamma_{yz}}^{\check p} F'_{x0\check p}
+ {\gamma_{zx}}^{\check p} F'_{y0\check p}
\right)
\right\}\Theta
\label{tdual}
\end{eqnarray}
which coincides with the quadratic action
\begin{eqnarray}
S_{WZ} &=& \mu_3\int d^4\zeta \sqrt{{\rm det} g} \epsilon^{\tilde i\tilde j\tilde k\tilde l}\left(
{1\over 4!\cdot 48} \bar\Theta{\Gamma_{\tilde i\tilde j\tilde k\tilde l}}^{pqr}\Theta F'_{pqr}
+ {1\over 48} \bar\Theta\Gamma_{\tilde i\tilde j\tilde k}\Theta F_{\tilde l}
\right. \cr & + & \left.
 {1\over 32} \bar\Theta{\Gamma_{\tilde i\tilde j}}^p\Theta F'_{\tilde k\tilde lp}
- {1\over 16\cdot 3!} \bar\Theta{\Gamma_{\tilde i}}^{pq}\Theta F'_{\tilde j\tilde k\tilde lpq}
\right)
\end{eqnarray}
We have chosen the gauge, Eq.(\ref{theta}), in constructing the D3 brane action. Agreement with 
the D0 brane action shows that this agrees with the gauge choice, 
$\Gamma^{11}\Theta=-\Theta$ in the IIA case. This point was already noted in \cite{Grana}. 

As was mentioned in the introduction the action for branes upto quadratic order in fermions 
in the presence of  an arbitary on-shell background has already been derived 
by Marolf, Martucci and Silva \cite{Marolf1}, \cite{Marolf2}.  
These authors used the method of ``normal coordinate expansion'' together with T-duality which is 
 different from the method of gauge completion used here. As we discuss below  our results completely agree. 
This constitutes a significant check of our results and methods. 

The quadratic fermion terms in the action for a $D_p$ brane are given in eq.(30) of \cite{Marolf2}.
We are interested in the case $p=3$ here. 
${\tilde \Gamma}_{D_p}$ is defined in eq.(28)   and  $L_p$ in eq.(29) of \cite{Marolf2}, with 
$\Gamma^\phi=-\sigma_3$ in our
notation.  Also,  $D_m$ and $\Delta$ are  defined in eq.(84), (86) of \cite{Marolf2}.
$y$ in eq.(30) of \cite{Marolf2}
 stands for the $32$ component spinor that we call $\theta$, with $y_1,y_2$ corresponding to
$\theta_1,\theta_2$ respectively . 
Let us for simplicity now set the world volume magnetic field to zero.
In the gauge $y_2=0$, it is then easy to see that  eq.(30) of \cite{Marolf2}
 agrees completely with the fermion bilinear terms obtained above, eq.(\ref{sdbi}), eq.(\ref{wzcs}),  
after identifying 
$y_1$ with $\Theta$ and the RR field strengths with each other upto a sign.

Finally,  the world volume action of $M5$ brane in presence of background flux
has been constructed by Kallosh and Sorokin \cite{Kallosh}. After a duality map this can be 
related to the D3 brane action computed here. We have compared with  the fermion bilinear terms
presented in eq.(22) of \cite{Kallosh} and find substantial agrement \footnote{We are greatful to R. Kallosh and
D. Sorokin for help in this regards.}.

\section{An Example}

\subsection{Euclidean Continuation}

The discussion above was for a D3 brane in  Minkowski space with signature $(9,1)$.
Our main interest here is in  instanton corrections to the superpotential 
and for this purpose we are really interested in   Euclidean space with signature $(10,0)$. 
We will not consider time dependent backgrounds here and continuing the bosonic fields which appear in the 
world volume theory Eqs.(\ref{sdbi}) and (\ref{wzcs}), to Euclidean space 
is straightforward. The world volume theory also contains a $16$ component Majorana
Weyl fermion, $\Theta$. This is continued to a $16$ component complex Weyl fermion in Euclidean space \footnote{
Note that there are no Majorana Weyl representations of $SO(10)$.}.
Fermion bilinear terms of the form:
\beq
\label{contbl}
S=\int d^4\xi \Theta^{T} \Gamma^0 M \Theta,
\eeq
are  continued to Euclidean space by replacing $\Theta$ above by the Weyl fermion. 
The path integral of the world-volume theory is then carried out over  $\Theta$ alone.
In particular  $\Theta^{\dagger}$ does not appear in the path integral  as an independent degree of freedom.  
In this way no further doubling of the fermionic degrees is introduced for the purposes of evaluating the 
path integral \cite{affleck}. 
It is also worth emphasising that on  analytic continuation to Euclidean space the
WZ term eq.(\ref{newwz}) picks up a factor of $i$ in front from the  continuation of the 
$\epsilon^{{\tilde i}{\tilde j} {\tilde k} {\tilde l}}$ symbol \footnote{This point was not correctly taken into account in earlier versions of the paper. We thank E. Bergshoeff, R. Kallosh, A. Kashani-Poor, D. Sorokin and A. Tomasiello
for pointing this out to us.}.

\subsection{The Example}

Now let us consider a specific example that will illustrate the role that fluxes can play in changing the count of 
zero modes. We consider a $T^6/Z_2$ compactification with flux \cite{kst,frey}.
 The six coordinates of torus are taken to be,
$x^i, y^i, i=1, \cdots 3$, with $0\le x^i,y^i\le 1$. The $Z_2$ orientifold symmetry involves
a reflection in all six directions, $(x^i,y^i) \rightarrow -(x^i,y^i), i=1, \cdots 3$.
 Holomorphic coordinates are,
$Z^i=x^i+\tau_{ij}y^j$, where $\tau_{ij}$ determine the complex structure of the torus. 
The tree-level superpotential takes the form, \cite{gukov,GKP},
\beq
\label{gvw}
W_{tree}=\int (F-\Phi H) \wedge \Omega_3
\eeq
where $\Phi=C+ie^{-\phi}$ is the axion-dilaton, and $\Omega_3$ is the holomorphic three-form which in this case takes the form, $\Omega_3=dZ^1\wedge dZ^2\wedge dZ^3$.

We focus on one specific choice of flux:
$F$ and $H$:
\begin{eqnarray}
 F &=& dx^1\wedge dx^2\wedge dx^3 + dy^1\wedge dy^2\wedge dy^3 \cr
 H &=& dx^1\wedge dx^2\wedge dx^3 - 2  dy^1\wedge dy^2\wedge dy^3
- dx^2\wedge dx^3\wedge dy^1 -  dx^3\wedge dx^1\wedge dy^2 \cr
&-& dx^1\wedge dx^2\wedge dy^3
+dy^2\wedge dy^3\wedge dx^1+dy^3\wedge dy^1\wedge dx^2+dy^1\wedge dy^2\wedge dx^3
\label{flux}
\end{eqnarray}
This example was analysed in \cite{kst} and it was shown that as a result of the superpotential,
Eq.(\ref{gvw}), all the 
complex structure moduli of the torus as well as  the axion-dilaton are stabilized with a value
\begin{eqnarray}
C + i e^{-\phi} = e^{2\pi i\over 3} ~, 
~~ \tau_{ij} = \delta_{ij} e^{2\pi i\over 3}.
\end{eqnarray}
The supersymmetry is broken to ${\cal N}=1$ in the resulting vacuum.

We are interested in possible non-perturbative corrections to the superpotential that can arise in this
${\cal N}=1$ theory.  Such corrections could arise due to Euclidean D3 branes wrapping divisor in $T^6/Z_2$.
A correction to the superpotential requires two fermionic zero  modes, no more or less, 
 in the world volume theory  of 
the Euclidean D3 brane.  Without flux there are $16$ fermionic zero modes. This is too many 
 (the sixteen zero modes follow
 from the ${\cal N}=4$ supersymmetry, present without flux, which also precludes a 
correction to the superpotential). With flux we will see below that ten 
zero modes survive.  
 This is fewer in number,
but still too many for a non-perturbative contribution to the superpotential.

A general divisor takes the form, $n_iZ^i=c$, where $n_i$ are integers
and $c$ is a constant.
We  first examine the  divisor $Z^3= c$ below. In 
this case the D3 wraps the $x^1,x^2,y^1,y^2,$ 
directions with $x^3,y^3,$ held constant. 
For now we also   exclude the special values,  $c=0, 1/2i, 1/2, 1/2+1/2i$. 
At these special values the $Z_2$ orientifold symmetry relates points on the divisor to each other.
This complicates the analysis somewhat. 
Towards the end of the section we  will consider the more general  divisor.
Using the symmetries of the problem we will find that the analysis can be mapped 
to the case when $Z^3=c$, thus resulting in the same number of zero modes.

We  ignore the five-form flux also   we set $F-B$ on the world volume to zero \footnote{
For the flux, Eq.(\ref{flux}), we can work in a gauge where the two-form RR gauge 
potential $C_{(2)}$ has non-zero components,
$C_{(2)x^1x^3}$, $C_{(2)y^1y^3}$. Since the brane extends along, $x^1,x^2,y^1,y^2,$ there is
then  no source term
for $F-B$ on the world volume and setting it to zero  is  consistent with the  equations of motion for the 
world volume gauge field.}. 
The fermion bilinear term of the action then takes the
 form
\begin{eqnarray}
S(\Theta^2) &=& - \mu_3\int d^4\zeta e^{-\phi} \sqrt{{\rm det}~g}
\left({1\over 48} \bar\Theta\Gamma^{mnp}\Theta H_{mnp}
- {1\over 16} \bar\Theta\Gamma_{\tilde ipq}\Theta H^{\tilde ipq}
\right) \cr
&+& i {\mu_3\over 32} \int d^4\zeta \epsilon^{\tilde i\tilde j\tilde k\tilde l}\left(
{1\over 36} \bar\Theta{\Gamma_{\tilde i\tilde j\tilde k\tilde l}}^{pqr}\Theta F'_{pqr}
+\bar\Theta{\Gamma_{\tilde i\tilde j}}^p\Theta F'_{\tilde k\tilde lp}
%-{1\over 3} \bar\Theta{\Gamma_{\tilde i}}^{pq}\Theta F'_{\tilde j\tilde k\tilde lpq}
\right) ~.
\label{bilin}
\end{eqnarray}
In this equation $\Theta$ is a Weyl fermion of $SO(10)$ but ${\bar \Theta}$ actually stands for 
$\Theta^T \gamma^0$, as was explained above.
We see that the flux gives rise to mass terms for the fermion $\Theta$.

The flux, Eq.(\ref{flux}) does not fix all the Kahler moduli.
With the choice,
\beq
\label{km}
ds^2=\sum_{a=1}^3 r_a^2 dz^ad{\bar z}^a
\eeq
it is easy to see that the Kahler moduli $r_a^2$,  contribute an overall multiplicative 
factor to the mass terms above.
Since our main goal is to count the zero modes here, we will work with $r_a=1$ below.
 
Now let us write the mass terms above as,
\begin{equation}
S(\Theta^2) = {\mu_3\over 8} \int d^4\xi \sqrt{{\rm det}g} \bar\Theta M \Theta ~,
\label{massm}
\end{equation}
where the matrix $M$ is determined by the flux. 
We are interested in the number of zero modes of $M$. 

As we discuss in Appendix {\bf A.6}, it is convenient to work in the following  basis
for the analysis. 
  Label the $16$ components of $\Theta$ as 
$|\epsilon_1, \epsilon_2,\epsilon_3, a>$ where $\epsilon_i=\pm 1, i=1, \cdots 3$
refer to the eigenvalues of $\Gamma^{\hat x^j\hat y^j}$ respectively{\footnote{
Here the `$~\hat{}~$' indicates that we are in the vierbein basis.}}.  E.g.,
\beq
\label{exeps}
\Gamma^{\hat x^1\hat y^1}|\epsilon_1,\epsilon_2,\epsilon_3,a>
= i \epsilon_1 |\epsilon_1,\epsilon_2,\epsilon_3,a>.
\eeq
And 
$a=\pm 1$ is an extra label (The $SO(10)$ rotation group  has a $SO(4) \times SO(6)$ subgroup where the $SO(6)$ 
refers to the compactified directions. The  label $a$ refers
to  the $SO(4)$, it takes only two values because the ten dimensional chirality  is fixed.).
Now it is easy to see that $M$ acts on the state, $|\epsilon_1,\epsilon_2,\epsilon_3,a>$, as follows,
\beq
\label{actms}
M |\epsilon_1,\epsilon_2,\epsilon_3,a>=\left({2\over \sqrt{3}}\right)^3  m ~~ \Gamma^{{\hat y}^1{\hat y}^2{\hat y}^3}
|\epsilon_1,\epsilon_2,\epsilon_3,a>,
\eeq
where,
\begin{eqnarray}
\label{deflm}
 m &=& 
  %\left\{\sin\left({2\pi\over 3}\right)
%- \cos\left({2\pi\over 3}\right) \epsilon_1\epsilon_2\right\}
{1\over 2} \left(\sqrt{3} + i \epsilon_1\epsilon_2\right)
\left\{e^{- {i\pi\over 3}(\epsilon_1+\epsilon_2+\epsilon_3)} - 2
-e^{- {i\pi\over 3}(\epsilon_1+\epsilon_2)}
-e^{- {i\pi\over 3}(\epsilon_2+\epsilon_3)} 
\right. \cr &-& \left.
e^{- {i\pi\over 3}(\epsilon_3+\epsilon_1)}
+ e^{-{i\pi\over 3}\epsilon_1}
+ e^{-{i\pi\over 3}\epsilon_2}
+ e^{-{i\pi\over 3}\epsilon_3}
\right\}
+ i \epsilon_1\epsilon_2
\left(1 + e^{-{i\pi\over 3}(\epsilon_1+\epsilon_2+\epsilon_3)}\right) ~.
\end{eqnarray}
Our notation for the matrix $\Gamma^{\hat y^1\hat y^2\hat y^3}$ is defined in Appendix {\bf A.6}.
We note here that $(\Gamma^{\hat y^1\hat y^2\hat y^3})^2=-1$ and thus 
$\Gamma^{{\hat y}^1{\hat y}^2{\hat y}^3} |\epsilon_1,\epsilon_2,\epsilon_3,a>$ cannot vanish.
This means that  the rhs of eq.(\ref{actms}) can vanish only if $m$ vanishes. 

It is easy to see from eq.(\ref{deflm}) that this happens when 
when $\epsilon_1=\epsilon_2=\epsilon_3=1$ or when $\epsilon_3 =-1, \epsilon_1 = \pm 1, \epsilon_2 = \pm 1 $. 
As discussed in Appendix A.6, 
these are the choices of $\epsilon_1, \epsilon_2,\epsilon_3$ for which $ m$ vanishes. 
Since $a$ in addition
can take values $\pm 1$, we get ten zero modes in all. 

This example illustrates the fact that fluxes can lift zero modes, although in this case
we see that the remaining number is still too large for a contribution to the
 superpotential.

\subsection{Discussion}
The analysis of zero modes in \cite{witten} cannot be directly applied 
 to the example above,
since the M-theory lift of the $T^6/Z_2$ orientifold is a space of reduced holonomy. 
Still, an analogous index can be defined in this example. The $U(1)$ symmetry
here  corresponds to rotations in the plane formed by  the $x^3,y^3$ directions. The $U(1)$ 
charge  of a zero mode is therefore simply $\epsilon_3$. The graded index is then,
\beq
\label{exind}
\chi\equiv \sum (-1)^{\epsilon_3},
\eeq
where the sum is over all the fermionic zero-modes. 
In the absence of flux, there are $8$ zero modes with $\epsilon_3=+1$ and 
$8$ with $\epsilon_3=-1$ so this index vanishes. 

We see from Eq.(\ref{flux}) that the three-form fluxes $H,F$ have two legs along the divisor and one normal to it, 
and so break the $U(1)$ symmetry. From the above discussion of the number of surviving zero modes we see that 
after the flux is turned on $\chi=-3$. 
In the more generic case of a Calabi-Yau space with flux also 
one expects that the index can change after incorporating flux. 
Evidence for this was already  found in 
\cite{GKTT}  for the case of M theory on  $K3\times K3$. There  it was argued 
that for a divisor of the form $K3\times P^1$, zero modes coming from $h(2,0)$ of the 
divisor, which have the same $U(1)$ charge,  pair up amongst themselves and get heavy. 

After turning on the flux, Eq.(\ref{flux}), ${\cal N}=1$ supersymmetry is left unbroken
in the resulting vacuum. The D3 brane wrapping the divisor breaks some of these supersymmetries,
and this gives rise to some  zero modes  in the D3 brane world volume theory.  
It would be helpful to know which of the  zero modes we have found above are related to the breaking 
of supersymmetry. We have not analysed this question in detail and leave it for the future. Let us note in 
passing here that  the conditions for supersymmetry imposed by the D3 brane are independent 
of the three-form flux. In the absence of flux, half the supersymmetries are left unbroken by the D3 brane wrapping a
 divisor, this suggests that two of the four supersymmetries are broken by the D3 brane, and two of the ten 
zero modes are due to this partial breaking of supersymmetry \footnote{We are grateful to Rudra Jena for a discussion 
in this regard.}.

We have focussed on a specific divisor above, $Z^3=c$.  
The case when $Z^3$ is replaced by $Z^1,Z^2$ gives the same zero-mode count due to 
the symmetries of the flux, Eq.(\ref{flux}).
Also in the discussion above we have excluded some  special values,   $c=0,1/2i, 1/2, 1/2(1+i)$. 
The divisors for these values of $c$  are special.  The $Z_2$ orientifold symmetry relates points on the 
divisor to each other in these cases  so the divisors  are ``half-cycles". 
Starting with a situation where the brane is away from one of these special values of $c$ we can 
continuously  move it to the special values. The brane  and its image under the $Z_2$ orientifold
 symmetry come together then.  Since the brane  can be moved continuously 
 in this way we do not expect the number of zero modes to jump.
A more interesting possibility is that of a 
brane without its image wraping one of these special divisors. 
This would be  the analogue of a fractional brane.
We have not fully explored this interesting case  yet and hope to 
return to it in the future.

A more general divisor has  the form, $n_iZ^i=c$. As  discussed towards  the end of Appendix
{\bf A.6}, upto an overall rescaling of the mass matrix, the analysis for the more general 
divisor  can be mapped to the case where one of the three coordinates, $Z^1,Z^2$ or $Z^3$ is a
 constant. Thus 
the discussion above applies and we 
learn  (again with the possible exception of some  special  values of $c$) that 
 for the case of a more general divisor   as well there
 are four fermion zero modes.  

Finally, the  zero modes we have found are  constant spinors which are zero modes of the mass matrix, 
eq.(\ref{massm}). 
They are therefore zero modes of the Dirac operator,
\begin{equation}
\label{do}
\Dslash \Theta + {1\over 4}M \Theta=0, 
\end{equation}
since both term above vanish seperately acting on the zero modes. 
One could ask if there are additional non-constant zero modes of the Dirac operator \footnote{We thank Renata
Kallosh for a discussion of this issue.}.
Under a rescaling of the volume of the internal space, $g_{mn} \rightarrow 
\lambda^2 g_{mn}$ (where now $m,n$ take values only over
 the six internal space directions) one finds that $\Dslash \rightarrow {1\over \lambda} \Dslash$ while
$M \rightarrow {1\over \lambda^3} M$. Thus at large volume the first term, $\Dslash \Theta$, is much more important
and our approximation of starting with constant spinors and seeking zero modes of $M$ amongst them is justified. 
Additional non-constant zero modes of the Dirac operator eq.(\ref{do})
 can be found  in this example, but in agreement with the argument just mentioned they always occur 
at a volume  of order the string scale. For such small volumes the $\alpha'$ corrections are important 
and the analysis is not trustworthy \footnote{In other examples
of a Calabi-Yau space with large orientifold charge the flux can be bigger and it might be possible to have the two 
terms in eq.(\ref{do}) comparable to each other when the volume is bigger than the string  scale.}.

\section{Conclusions}
In this paper we have used the method of gauge completion and 
determined the  fermion bilinear terms in the world volume action of
 a D3 brane in the presence of background flux. Our results are summarised in Eq.(\ref{sdbi}) and Eq.(\ref{wzcs}). 
These results have been previously obtained by Marolf, Martucci and Silva using somewhat different methods. 

The fermion bilinear terms  are of interest in calculating  instanton corrections to the superpotential
in flux compactifications.  They are also of interest in  determining 
soft susy breaking terms that can arise in flux compactifications.  

For a Euclidean D3 brane wrapping a divisor in a six dimensional compactification
these results explicitly show that 
 the $U(1)$ symmetry of rotations normal to the divisor is broken in the presence of three-form flux.
In an explcit example of a $T^6/Z_2$ compactification with three-form flux we have calculated the fermion mass terms 
and shown that some  zero modes are lifted due to the flux. 
  
There are several directions of future work. 
One would like a better understanding of supersymmetry in this context. 
This is connected   to the number of zero modes in the world volume of the D3 brane. 
More generally, one would like to use our  results to calculate the instanton corrections in situations where 
they can arise. Even in the simple example studied here, of a $T^6/Z_2$ compactification, our analysis is not complete
and the case when the $D3$ brane wraps a half-cycle needs to be understood better. 

We hope to return to these questions in the future.

\begin{center}
\large{\bf Acknowledgements}
\end{center}
We are grateful to Lars Goerlich for collaboration in the initial stages of this project.
We also acknowledge discussions with  Gabriel Cardoso, Gottfried Curio, Atish Dabholkar, Rudra Jena, 
Shamit Kachru, Renata Kallosh, Dieter L\"ust, Liam McAllister, Sunil Mukhi, Dmitri Sorokin 
and Angel Uranga.  
The work of PKT is supported by German Research Foundation (DFG) 
Project LU~419/7-3. The work of SPT is supported by a Swarnajayanti Fellowship,
DST, Government of India. We are deeply grateful  to  the people of India for 
enthusiastically supporting research in String Theory.

\newpage

\noindent{\large\bf Appendix}

\noindent
Our conventions are as follows. 

Superspace coordinates are denoted  by $Z^M = (x^m,\theta^\mu)$, which stand for the
bosonic and fermionic components respectively.
Curved space indices are given by  
$\{M,N,\cdots\}=\{m,n,\cdots,\mu,\nu,\cdots\}$ 
where $(m,n)$ denote Bosonic indices and $(\mu, \nu)$ fermionic indices.
Tangent space indices are given by
$\{A,B,\cdots\}=\{a,b,\cdots,\alpha,\beta,\cdots\}$, with $(a,b)$ denoting bosonic and
$(\alpha,\beta)$ fermionic indices. 

We will use real $16$ component Majorana-Weyl spinors, a convenient basis of 
gamma matrices is given in Eq.(4.3.48), \cite{GSWa}, in which 
$\Gamma^0$ is antisymmetric and the remaining gamma matrices,
$\Gamma^i$, are symmetric. Since there are two  $16$ component Majorana-Weyl spinors 
worth of supersymmetries in the IIB theory our spinors will carry an extra $SO(2)$ index.
The spinor indices $\alpha,\beta,$ should be viewed as composite indices standing for the
tensor product of a Majorana-Weyl index and this  additional  $SO(2)$ index.
In the formulae below the gamma matrices will act on the Majorana Weyl index while the 
Pauli spin matrices, $\sigma^1, \sigma^2,\sigma^3,$ will act on the $SO(2)$ index.

Throughout this paper, we  denote  antisymmetrisation with unit weight by a 
square  bracket. For example, the antisymmetrised product of an antisymmetric
  rank-two tensor $A_{mn}$  with 
a rank one tensor $B_p$ is,  
\beq
\label{antisycon}
 A_{[mn} B_{p]}={1\over 3}[A_{mn}B_{p}-A_{pn}B_{m} -A_{mp}B_{n}].
\eeq
There are $3$ distinct terms which appear on the rhs as shown.
To make it of unit weight we divide by the number of distinct terms, which accounts for the 
prefactor  ${1\over 3}$.
Finally, 
$\Gamma_{m_1\cdots m_n} = \Gamma_{[m_1}\cdots\Gamma_{m_n]}$,
will denote the  antisymmetrised product of $n$ Gamma matrices. 

\noindent
{\large\bf A.1 Supersymmetry Transformations}

With these conventions, the supersymmetric transformation rules in the 
string frame \cite{deroo} 
are given by
\footnote{Note that here
our  normalization for $\lambda$ is different from Ref.\cite{deroo}}
\begin{eqnarray}
&& \delta\lambda = {1\over 2} \Gamma^m\partial_m\phi\epsilon
- {1\over 24} \Gamma^{mnp} H_{mnp} \sigma^3\epsilon
- {1\over 2} e^\phi \Gamma^m F_m (i\sigma^2)\epsilon
- {1\over 24} e^\phi \Gamma^{mnp} F'_{mnp} \sigma^1\epsilon \cr
&& \delta\psi_m = D_m\epsilon 
+ {1\over 8} e^\phi \Gamma^p\Gamma_m F_p (i\sigma^2)\epsilon
- {1\over 8}\Gamma^{pq} H_{mpq} \sigma^3\epsilon
+ {1\over 48} e^\phi \Gamma^{pqr} \Gamma_m F'_{pqr} \sigma^1\epsilon \cr
&& ~~~~~~~~~~~~~ + {1\over 16\cdot 5!} e^\phi
\Gamma^{pqrst} \Gamma_m F'_{pqrst} (i\sigma^2)\epsilon \cr
&& \delta\phi = \bar\epsilon\lambda \cr
&& \delta C = e^{-\phi} \bar\epsilon\sigma^1\lambda \cr
&& \delta{e_m}^a = \bar\epsilon\Gamma^a\psi_m \cr
&& \delta{B}_{mn} =  \bar\epsilon\sigma^3
\left(\Gamma_m\psi_n - \Gamma_n\psi_m\right) \cr
&& \delta{C}_{mn} = e^{-\phi}\bar\epsilon\sigma^1\left(\Gamma_{mn}\lambda
- 2 \Gamma_{[m}\psi_{n]}\right) + C \delta{B}_{mn} \cr
&& \delta{C}_{mnpq} = e^{-\phi}\bar\epsilon (i\sigma^2) \left(
\Gamma_{mnpq}\lambda - 4 \Gamma_{[mnp}\psi_{q]}\right)
+ 6 C_{[mn}\delta{B}_{pq]} ~.
\label{susytr}
\end{eqnarray} 
Here $F'_3$ and $F'_5$ are the gauge invariant RR field strengths
\begin{equation}
F'_3 = dC_2 - C H_3 ~, ~~ F'_5 = dC_4 - C_2\wedge H_3 ~.
\end{equation}
Using these supersymmetry transformations, in the following sections
we will compute the the expansion of the superfields 
$\hat C_{MN}, \hat C_{MNPQ}$ and ${\hat e_M}^A$,
up to $O(\theta^2)$, in terms of the component fields. 

\noindent
{\large{\bf A.2 ~~ Calculation for} $\hat C_{MN}$}

Following similar steps for the calculation of ${\hat B}_{MN}$ in \S2.2, 
here we will carry out the expansion of $\hat{C}_{MN}$ to $O(\theta^2)$. 
For this purpose, we must supply the superspace gauge transformation 
$\Sigma_M^{(c)}$ to $O(\theta)$, in addition to the super diffeomorphism 
(\ref{sdotheta}). 

Let us first evaluate the commutator of two supersymmetry transformations
(with parameters $\epsilon^1,\epsilon^2$) on the field $C_{mn}$.
%$$(\delta_1\delta_2 - \delta_2\delta_1) C_{mn}.$$ Using the susy 
Using Eq.(\ref{susytr}) for the supersymmetry transformation of $C_{mn}$
we find that 
\begin{eqnarray}
\delta_1\delta_2 C_{mn} = 
e^{-\phi}\bar\epsilon_2\left(
\Gamma_{mn}\delta_1\lambda - 2 \Gamma_{[m}\delta_1\psi_{n]}\right)
+ C \delta_1\delta_2 B_{mn} ~.
\end{eqnarray}
Using Eq.(\ref{susytr}) once more, it is straightforward to see that
the commutator can be expressed as a diffeomorphism (with the parameter
$\xi^m$ as given by Eq.(\ref{sdotheta})), and a gauge transformation 
$\xi^{(c)}_m$. In other words
\begin{eqnarray}
(\delta_1\delta_2 - \delta_2\delta_1) C_{mn} 
= \xi^p\partial_p C_{mn} + \partial_m\xi^p C_{pn} - \partial_n\xi^p C_{pm}
+ \partial_m\xi_n^{(c)} - \partial_n\xi_m^{(c)} ~,
\label{commcmn}
\end{eqnarray}
with the gauge transformation parameter
\begin{eqnarray}
\xi_m^{(c)} &=& \xi^n C_{mn}
+ e^{-\phi} \bar\epsilon_2\sigma^1\Gamma_m\epsilon_1
- C \bar\epsilon_2\sigma^3\Gamma_m\epsilon_1 ~.
\end{eqnarray}
In order to find the gauge transformation parameter, we have to compare 
Eq.(\ref{commcmn}) with the commutator derived in the superspace formalism. 
It is easy to see that 
\begin{eqnarray}
(\delta_1\delta_2 - \delta_2\delta_1){\hat C}_{MN}
= \left(\partial_M\Sigma_N^{12(c)} 
- (-1)^{MN} \partial_N\Sigma_N^{12(c)}\right) + \cdots ~,
\label{conetwo}
\end{eqnarray}
where the dots denote terms arising due to superdiffeomorphism. 
The superspace gauge transformation parameter $\Sigma^{12(c)}_M$ is given by
\begin{eqnarray}
\Sigma_M^{12(c)} = \left(\Sigma_2^P\partial_P\Sigma^{1(c)}_M
+ \partial_M\Sigma_2^P \Sigma^{1(c)}_P\right)
- \left(1\leftrightarrow 2\right) ~.
\label{cxiontw}
\end{eqnarray}
The commutator on component field $C_{mn}$ will agree with the commutator 
on the superfield $\hat{C}_{mn}$ if the gauge transformation parameter 
$\Sigma^{(c)}_m$ takes the value
\begin{eqnarray}
\Sigma_m^{(c)} &=& {1\over 2} \bar\theta\Gamma^n\epsilon C_{mn}
+ {1\over 2} \bar\theta\left(e^{-\phi}\sigma^1
- C\sigma^3\right)\Gamma_m\epsilon ~.
\label{gaugetrc}
\end{eqnarray}
The component fields $C_{m\mu}$ and $C_{\mu\nu}$ are both zero to leading
order and hence the commutator of the two susy on them also vanishes. From
this it is easy to see that the component $\Sigma^{(c)}_\mu$ vanishes
for the case when the the space time fermion backgrounds are set to zero.

Now let us compute the first order expansion for ${\hat C}_{mn}$. Comparing
the susy transformation for $C_{mn}$ from Eq.(\ref{susytr}) with the 
superfield result 
$\delta\hat C_{mn} = \epsilon^\alpha\partial_\alpha \hat C_{mn}$, we find 
\begin{eqnarray}
{\hat C}_{mn} = C_{mn} + e^{-\phi}\bar\theta\sigma^1\left(\Gamma_{mn}\lambda
- 2 \Gamma_{[m}\psi_{n]}\right) 
+ 2 C \bar\theta \Gamma_{[m}\psi_{n]} ~.
\label{cmnot}
\end{eqnarray}
The expression for $\hat C_{m\mu}$ can similarly be derived. 
Using the expression
for the gauge transformation (\ref{gaugetrc}) the superspace variation for 
$\hat C_{m\mu}$ can be written as 
\begin{eqnarray}
\delta{\hat C}_{m\mu} =  \epsilon^\alpha\partial_\alpha{\hat C}_{m\mu}
- {1\over 2} e^{-\phi} \left(\bar\epsilon\sigma^1\Gamma_m\right)_\mu
+ {1\over 2} C \left(\bar\epsilon\sigma^3\Gamma_m\right)_\mu ~.
\label{rhscmu}
\end{eqnarray}
Since the component field susy transformation $\delta C_{m\mu} = 0$, 
the r.h.s. of Eq.(\ref{rhscmu}) has to be equated to zero. This gives
the expression
\begin{eqnarray}
\hat{C}_{m\mu}
= {1\over 2} e^{-\phi} \left(\bar\theta\sigma^1\Gamma_m\right)_\mu
  - {1\over 2} C \left(\bar\theta\sigma^3\Gamma_m\right)_\mu  ~.
\end{eqnarray}
With the help of this equation and the gauge transformation (\ref{gaugetrc}),
we can write down the variation of the superfield $\hat C_{mn}$ up to 
$O(\theta)$.
\begin{eqnarray}
\delta{\hat C}_{mn} &=& \epsilon^\alpha\partial_\alpha{\hat C}_{mn}
- 2 \bar\theta\left( e^{-\phi}\sigma^1
- C \sigma^3\right) \Gamma_{[m}\partial_{n]}\epsilon 
+ \bar\theta\left( e^{-\phi}\sigma^1
- C \sigma^3\right)\Gamma^a\epsilon \partial_{[m}e_{n]a} \cr
&-& {1\over 2}  \bar\epsilon\Gamma^q\theta F_{qmn} 
+ \left(e^{-\phi}\bar\epsilon\sigma^1\Gamma_{[m}\theta\partial_{n]}\phi
+ \bar\epsilon\sigma^3\Gamma_{[m}\theta\partial_{n]}C\right) ~.
\end{eqnarray}
On the other hand, we can use Eq.(\ref{cmnot}) for $\hat C_{mn}$ to arrive at
\begin{eqnarray}
\delta{\hat C}_{mn} =\delta C_{mn} 
+ e^{-\phi}\bar\theta\sigma^1\left(\Gamma_{mn}\delta\lambda
- 2 \Gamma_{[m}\delta\psi_{n]}\right) 
+ 2 C \bar\theta \Gamma_{[m}\delta\psi_{n]} ~.
\end{eqnarray}
These two variations must be the same. 
When we plug in the susy transformations for $\psi_m$ and $\lambda$ from 
Eq.(\ref{susytr}), we find that they will match up only when  $\hat C_{mn}$ 
has the following expression to second order in $\theta$:
\begin{eqnarray}
{\hat C}_{mn} &=& C_{mn} + e^{-\phi}\bar\theta\sigma^1\left(\Gamma_{mn}\lambda
- 2 \Gamma_{[m}\psi_{n]}\right)
+ 2 C \bar\theta \Gamma_{[m}\psi_{n]} 
+ {1\over 4} e^{-\phi} \bar\theta\sigma^1\Gamma_{mnp}\theta\partial^p\phi \cr 
&+& {1\over 8} \bar\theta\left( e^{-\phi}\sigma^1 - C \sigma^3\right)
\left({\Gamma_m}^{ab}\omega_{nab} - {\Gamma_n}^{ab}\omega_{mab}\right)\theta
+ {1\over 8}\bar\theta
\left(\sigma^3 - e^{\phi} C\sigma^1\right)\Gamma_{mnp}\theta F^p \cr 
&+& {1\over 8} e^{-\phi}\bar\theta(i\sigma^2)\Gamma^p\theta H_{mnp} 
+ {1\over 48} e^{-\phi}\bar\theta(i\sigma^2){\Gamma_{mn}}^{pqr}\theta H_{pqr}
- {1\over 8} C \bar\theta{\Gamma_{[m}}^{pq} H_{n]pq}\theta \cr 
&-& {1\over 8}  \bar\theta{\Gamma_{[m}}^{pq} F'_{n]pq}\theta 
- {1\over 8} C e^{\phi} \bar\theta(i\sigma^2)\Gamma^p\theta F'_{mnp}
- {1\over 48} C e^\phi\bar\theta(i\sigma^2){\Gamma_{mn}}^{pqr}\theta F'_{pqr}\cr
&-&{1\over 16\cdot 5!} \bar\theta\left(\sigma^3 + C e^\phi \sigma^1\right)
\left({\Gamma_{mn}}^{pqrst} F'_{pqrst} + 20 \Gamma^{pqr} F'_{mnpqr}\right)
\theta ~.
\end{eqnarray}

\noindent
{\large{\bf A.3 ~~ Calculation for} $\hat C_{MNPQ}$} 

Let us now turn to $\hat{C}_{MNPQ}$.  The calculation is pretty much the same
as the previous ones for $\hat B_{MN}$ and $\hat C_{MN}$.  We will first 
evaluate the gauge transformation parameter $\Sigma_{MNP}$ from the commutator 
of two susy transformation on the component field $C_4$ and then use this 
information to derive the $O(\theta^2)$ expression for the superfield 
$\hat C_4$. From Eq.(\ref{susytr}) we find
\begin{eqnarray}
\delta_1\delta_2 C_{mnpq} = e^{-\phi}\bar\epsilon_2 (i\sigma^2)\left(
\Gamma_{mnpq}\delta_1\lambda - 4 \Gamma_{[mnp}\delta_1\psi_{q]}\right)
+ 12 \bar\epsilon_2\sigma^3 C_{[mn}\Gamma_p\delta_1\psi_{q]} ~.
\end{eqnarray}
After some straightforward calculation, the commutator of two susy 
transformations can be written as 
\begin{eqnarray}
(\delta_1\delta_2 - \delta_2\delta_1) {C}_{mnpq} &=&
\bar\epsilon_2\left(
- 4 e^{-\phi} (i\sigma^2)\partial_{[m}\phi \Gamma_{npq]}
+ \Gamma^a F'_{amnpq} + 4 \sigma^3 \Gamma_{[m} F'_{npq]} 
\right. \cr & + &  \left.
4 \sigma^1 \Gamma_{[m} H_{npq]} + 6 C_{[mn}\Gamma^a H_{pq]a} 
\right)\epsilon_1 ~.
\end{eqnarray}
This is equal to a  diffeomorphism with diffeomorphism parameter $\xi^m$ as
given in Eq.(\ref{susytrans}), and a gauge transformation
$$d\xi_3 + d(H_3\wedge\xi^{(c)}) ~,$$
with the gauge transformation parameter $\xi_3$ having the expression 
\begin{eqnarray}
\xi_{mnp} = \xi^q C_{mnpq} 
              + e^{-\phi}\bar\epsilon_2 (i\sigma^2)\Gamma_{mnp}\epsilon_1
              - 3 C_{[mn}\bar\epsilon_2\sigma^3\Gamma_{p]}\epsilon_1 ~.
\end{eqnarray}
Now we can evaluate the commutator on the super field $\hat C_4$,
\begin{eqnarray}
(\delta_1\delta_2 - \delta_2\delta_1)\hat C_4 = d\left(\Sigma^{12}_3\right) 
+ \cdots ~.
\end{eqnarray}
Here again the dots denote the superdiffeomorphisms. The gauge transformation
parameter $\Sigma^{12}_3$ can be written in terms of components as 
\begin{eqnarray}
\Sigma_{MNP}^{12} = \left[\left(\Sigma_2^Q\partial_Q\Sigma_{1MNP} 
+ 3 \partial_{[M} \Sigma_2^Q\Sigma_{1NP]Q}\right)
                         - (1\leftrightarrow  2)\right]  ~.
\end{eqnarray}
Comparing the two commutators we can easily solve for the gauge
transformation parameter $\Sigma_{mnp}$ to obtain 
\begin{eqnarray}
\Sigma_{mnp} =  {1\over 2} \bar\theta\Gamma^q\epsilon C_{mnpq} 
+ {1\over 2} e^{-\phi} \bar\theta(i\sigma^2)\Gamma_{mnp}\epsilon
- {3\over 2} C_{[mn} \bar\theta\sigma^3\Gamma_{p]}\epsilon  ~.
\label{smnp}
\end{eqnarray}
All the remaining components of $\Sigma_{MNP}$ will be zero.

We now need to evaluate the expressions for $\hat C_{mnpq}$ and 
$\hat C_{\mu mnp}$ to $O(\theta)$. It is easy to see that the variation
$\delta \hat C_{mnpq} = \epsilon^\alpha\partial_\alpha \hat C_{mnpq}$,
and the susy transformation for $C_{mnpq}$ from Eq.(\ref{susytr}) gives
\begin{eqnarray}
\hat C_{mnpq} &=& C_{mnpq} + e^{-\phi}\bar\theta(i\sigma^2)\left\{
\Gamma_{mnpq}\lambda - 4 \Gamma_{[mnp}\psi_{q]}\right\}
+ 12 \bar\theta\sigma^3 C_{[mn} \Gamma_p \psi_{q]} ~.
\label{cmnpq}
\end{eqnarray}
Using the expression for the gauge transformation (\ref{smnp}), we can obtain
the variation $\delta\hat C_{\mu mnp}$. Since $C_{\mu mnp}$ vanishes to leading
order, this variation has to be set to zero. As a result, we get
\begin{eqnarray}
\hat C_{\mu mnp} = 
- {1\over 2} e^{-\phi} \left(\bar\theta(i\sigma^2)\Gamma_{mnp}\right)_\mu
+ {3\over 2}\left( C_{[mn}\bar\theta\sigma^3\Gamma_{p]}\right)_\mu ~.
\end{eqnarray} 
It is easy to see that all the remaining components of $\hat C_{MNPQ}$ vanishes.
Now we are ready to execute the second order results.Using the above expression 
for $\hat C_{\mu mnp}$ and the expression for the gauge transformation parameter
from Eq.(\ref{smnp}) we find the variation of $\hat C_{mnpq}$ to be of the form
\begin{eqnarray}
\delta \hat C_{mnpq} &=& \epsilon^\alpha\partial_\alpha\hat C_{mnpq}
+ {1\over 2} \bar\theta\Gamma^a\epsilon F_{amnpq} 
- 4 H_{[mnp} \Sigma^{(c)}_{q]} 
+4 \partial_{[m}\epsilon^\alpha \hat C_{\alpha npq]}  
\cr
&& + 2 \partial_{[m}\left(
e^{-\phi}\bar\theta(i\sigma^2)\Gamma_{npq]}\epsilon\right)
- 6 \partial_{[m}\left(C_{np}\bar\theta\sigma^3\Gamma_{q]}\epsilon\right) ~.
\end{eqnarray}
After substituting the expression for $\Sigma^{(c)}_m$ and making 
some rearrangement we get
\begin{eqnarray}
&& \delta \hat C_{mnpq} = \epsilon^\alpha\partial_\alpha\hat C_{mnpq}
+ {1\over 2} \bar\theta\Gamma^a\epsilon ( F_{amnpq} - 4 H_{[mnp} C_{q]a})
+ 2 \bar\theta e^{-\phi}\sigma^1 \Gamma_{[m}\epsilon H_{npq]}\cr
&& + 4 \partial_{[m}\epsilon^\alpha \hat C_{\alpha npq]}
+ 2\bar\theta\left\{ \partial_{[m}\left(
e^{-\phi}(i\sigma^2)\Gamma_{npq]}\epsilon\right)
-3C_{[np}\partial_m\left(\sigma^3\Gamma_{q]}\epsilon\right)
-F'_{[mnp}\sigma^3\Gamma_{q]}\epsilon\right\}~.
\end{eqnarray}
We can also obtain the variation from Eq.(\ref{cmnpq}) for 
the expansion of $\hat C_{mnpq}$ to $O(\theta)$:
\begin{eqnarray}
\delta\hat C_{mnpq} = \delta C_{mnpq} 
+ e^{-\phi}\bar\theta(i\sigma^2)\left\{
\Gamma_{mnpq}\delta\lambda 
- 3 \Gamma_{[mnp}\delta\psi_{q]}\right\}
+ 12 \bar\theta\sigma^3 C_{[mn} \Gamma_p \delta\psi_{q]} ~.
\end{eqnarray}
These two expressions must agree. This can be used to solve for $\hat C_{mnpq}$
to second order in $\theta$ to obtain
\begin{eqnarray}
&&\hat{C}_{mnpq} = C_{mnpq}
+ e^{-\phi}\bar\theta(i\sigma^2)\left\{
\Gamma_{mnpq}\lambda - 4 \Gamma_{[mnp}\psi_{q]}\right\}
+  12 \bar\theta\sigma^3 C_{[mn} \Gamma_p \psi_{q]}  \cr
&&+ {1\over 2} e^{-\phi} \bar\theta(i\sigma^2)\Gamma_{ab[mnp}
{\omega_{q]}}^{ab}\theta 
+ 3 e^{-\phi}\bar\theta(i\sigma^2)\Gamma_{[p}\omega_{qmn]}\theta
+{1\over 4} e^{-\phi} \bar\theta(i\sigma^2) {\Gamma_{mnpq}}^s
\partial_s\phi \theta  \cr
&&+ {1\over 48} e^{-\phi} \bar\theta\sigma^1{\Gamma_{mnpq}}^{stu} H_{stu}\theta
+ {1\over 48} \bar\theta\sigma^3{\Gamma_{mnpq}}^{stu} F'_{stu}\theta 
+ {1\over 2} \bar\theta\Gamma_{[mnp} F_{q]}\theta  \cr 
&&+ {3\over 4}\bar\theta\sigma^3 {\Gamma_{[mn}}^s F'_{pq]s} \theta 
+ {3\over 4} e^{-\phi} \bar\theta \sigma^1{\Gamma_{[mn}}^s H_{pq]s}\theta
- {1\over 96}\bar\theta {\Gamma_{[mnp}}^{stuv} F'_{q]stuv}\theta \cr
&&-{1\over 8}\bar\theta{\Gamma_{[m}}^{st}F'_{npq]st}\theta  
-{3\over 2} \bar\theta\sigma^3
C_{[mn} {\Gamma_p}^{ab} \omega_{q]ab}\theta  
- {3\over 4} e^\phi C_{[mn}\bar\theta\left( 
\sigma^1 {\Gamma_{pq]}}^s F_s
+ e^{-\phi} {\Gamma_p}^{st} H_{q]st} 
\right. \cr && \left.
+ i\sigma^2\left\{\Gamma^s F'_{pq]s}+{1\over 6}{\Gamma_{pq]}}^{stu}
F'_{stu}\right\}
+ {1\over 12} \sigma^1 \left\{\Gamma^{stu} F'_{pq]stu}
+ {1\over 20} {\Gamma_{pq]}}^{stuvw} F'_{stuvw} \right\}
\right)\theta ~.
\end{eqnarray}

\noindent
{\large\bf A.4 ~~ The Supervierbein}

Finally we come to the computation of the vierbeins. A similar calculation
can be performed in this case also. Note that in addition to the 
superdiffeomorphism, here we have to consider the (super) local Lorentz
transformation (\ref{lorenttrans}). Let us first compute vierbeins 
to $O(\theta)$. Equating 
$\delta{\hat e_m}^a = \epsilon^\alpha\partial_\alpha {\hat e_m}^a$ 
with $\delta{e_m}^a = \bar\epsilon\Gamma^a\psi_m$ we find
\begin{eqnarray}
{\hat e_m}^a = {e_m}^a + \bar\theta\Gamma^a\psi_m ~.
\label{vierfst}
\end{eqnarray}
Similarly we can compute ${{\hat e}_\mu}^a$ to $O(\theta)$. 
using the value of $\Sigma^m$ from Eq.(\ref{sdotheta}) we find, 
\begin{eqnarray}
{{\hat e}_\mu}^a &=& - {1\over 2} \left(\bar\theta\Gamma^a\right)_\mu ~.
\end{eqnarray}
To obtain the Lorentz transformation parameter to $O(\theta)$, we need to 
compute the commutator of two supersymmetry transformations on the vierbein
${e_m}^a$. This can be easily computed using the susy transformations 
(\ref{susytr}). After some simplification, we get
\begin{eqnarray}
\left(\delta_2\delta_1 - \delta_1\delta_2\right) {e_m}^a
& = & \left( \bar\epsilon_1\Gamma^a\partial_m\epsilon_2
- \bar\epsilon_2\Gamma^a\partial_m\epsilon_1\right)
 - \bar\epsilon_1\Gamma_b\epsilon_2{\omega_m}^{ab}  %\cr
+ {1\over 4} e^\phi \bar\epsilon_1{\Gamma^{ap}}_m (i\sigma^2)\epsilon_2 F_p
\cr &-& {1\over 2} \bar\epsilon_1\Gamma_q\sigma^3\epsilon_2{H_m}^{aq} +
{1\over 24} e^\phi \bar\epsilon_1{\Gamma^{apqr}}_m\sigma^1\epsilon_2F'_{pqr}
+{1\over 4} e^\phi \bar\epsilon_1\Gamma_q\sigma^1\epsilon_2{F'_m}^{aq} \cr
 & + & {1\over 8\cdot 5!} e^\phi \bar\epsilon_1{\Gamma^{apqrst}}_m
(i\sigma^2)F'_{pqrst}\epsilon_2 
+ {1\over 48} e^\phi\bar\epsilon_1 \Gamma^{pqr}(i\sigma^2)\epsilon_2 F'_{mapqr}
~.
\end{eqnarray}
The above equation can be written in the following simple form 
\begin{eqnarray}
\left(\delta_1\delta_2 - \delta_2\delta_1\right) {e_m}^a
= \xi^n\partial_n {e_m}^a + (\partial_m \xi^n) {e_n}^a + \lambda^{ab} e_{mb} ~,
\label{vircomm}
\end{eqnarray}
provided the translation parameter $\xi^n$ is given in Eq.(\ref{susytrans}), 
and the rotation parameter $\lambda^{ab}$ has the expression
\begin{eqnarray}
 \lambda^{ab} &=& - \xi^n{\omega_n}^{ab} 
+ {1\over 2} \bar\epsilon_2\Gamma_p\sigma^3\epsilon_1 H^{abp} 
- {1\over 4} e^\phi\bar\epsilon_2\left\{
\Gamma^{abp}(i\sigma^2) F_p
+ {1\over 6} \Gamma^{abpqr}\sigma^1 F'_{pqr}
\right. \cr &+& \left. \Gamma_p\sigma^1 {F'}^{abp} 
+ {1\over 2\cdot 5!}\Gamma^{abpqrst}(i\sigma^2)F'_{pqrst}
+ {1\over 12} \Gamma_{pqr}(i\sigma^2) {F'}^{abpqr}\right\}\epsilon_1 ~.
\end{eqnarray}
In deriving (\ref{vircomm}) we have used the following identity obeyed by
the spin connection and the vierbein
\begin{eqnarray}
e_{nb} {\omega_m}^{ab} = e_{mb} {\omega_n}^{ab} 
                      + \left(\partial_m{e_n}^a - \partial_n{e_m}^a\right) ~.
\end{eqnarray}
On the other hand, one can apply the commutator directly on the super vierbein
as given in Eq.(\ref{vierfst}). This will be consistent with Eq.(\ref{vircomm})
if the parameter  $\Lambda^{ab}$ takes the form 
\begin{eqnarray}
\Lambda^{ab}(\epsilon) 
& = & - {1\over 2} \bar\theta\Gamma^n\epsilon{\omega_n}^{ab}
+ {1\over 4} \bar\theta\Gamma_p\sigma^3\epsilon H^{abp} 
- {1\over 8} e^\phi\bar\theta\left(\Gamma^{abp}(i\sigma^2) F_p
+ {1\over 6} \Gamma^{abpqr}\sigma^1 F'_{pqr}
\right. \cr &+& \left.
\Gamma_p\sigma^1 {F'}^{abp} 
+ {1\over 2\cdot 5!} e^\phi \Gamma^{abpqrst}(i\sigma^2) F'_{pqrst}
+ {1\over 12} \Gamma_{pqr}(i\sigma^2) {F'}^{abpqr} \right)\epsilon ~.
\end{eqnarray}
Now we are ready to compute the $O(\theta^2)$ part of the super vierbein.
Consider the variation
\begin{eqnarray}
\delta{{\hat e}_m}^a & = & \Sigma^P\partial_P{{\hat e}_m}^a 
+\partial_m\Sigma^P{{\hat e}_P}^a + \Lambda^{aP}{{\hat e}_{mP}} \cr
& = & \epsilon^\alpha\partial_\alpha{{\hat e}_m}^a 
+ {1\over 2} \bar\theta\Gamma^n\epsilon\left(\partial_n{{\hat e}_m}^a
- \partial_m{{\hat e}_n}^a\right)
+ \bar\theta\Gamma^a\partial_m\epsilon + \Lambda^{ab}{{\hat e}_{mb}} ~.
\end{eqnarray}
It should be equated with the variation coming from Eq.(\ref{vierfst}),
\begin{eqnarray}
\delta{{\hat e}_m}^a =  \delta{e_m}^a + \bar\theta\Gamma^a\delta\psi_m ~,
\end{eqnarray}
from which it follows that 
\begin{eqnarray}
\epsilon^\alpha\partial_\alpha{{\hat e}_m}^a 
= \bar\theta\Gamma^a\delta\psi'_m - \Lambda^{ab} e_{mb}
- {1\over 2} \bar\theta\Gamma^n\epsilon\left(\partial_n{{e}_m}^a
- \partial_m{{e}_n}^a\right) ~.
\end{eqnarray}
Here the prime indicates the absence of $\partial_m\epsilon$ from
susy variation of the gravitino. Again using the formula for the 
supersymmetry transformations (\ref{susytr}), the above expression 
can easily be integrated. The super vierbein, up to $O(\theta^2)$, 
takes the form
\begin{eqnarray}
 {\hat e_m}^a & = &  {e_m}^a + \bar\theta\Gamma^a\psi_m 
- {1\over 8} \omega_{mcd} \bar\theta \Gamma^{acd}\theta %\cr & + & 
+ {1\over 16} e^\phi \bar\theta (i\sigma^2)\left(\Gamma_m F^a + \Gamma^a F_m
- \delta_m^a \Gamma^p F_p\right)\theta \cr & - &
 {1\over 16} \bar\theta\Gamma^{apq} \sigma^3\theta H_{mpq}
+ {1\over 32} e^\phi \bar\theta\left(\Gamma^{apq} F'_{mpq}
+ \Gamma_{mpq} {F'}^{apq}\right)\sigma^1\theta \cr
&+& {1\over 32\cdot 4!}e^\phi\bar\theta\left(\Gamma^{apqrs} F'_{mpqrs} 
+ \Gamma_{mpqrs} {F'}^{apqrs}\right)(i\sigma^2)\theta ~.
\end{eqnarray}

\noindent
{\large\bf A.5 ~~ T-duality}

It is in fact possible to obtain the $D3$ brane action, starting with $D0$
brane action and performing three T-dualities (say, along $x,y,z$). For 
simplicity, we assume the metric to be diagonal along the directions on which 
we perform T-duality. Also we set $B_{xi}=g_{xi}=0$ (and similar relations
for $y$ and $z$ directions). Here we summarize the rules for T-duality along 
the direction $x$. See \cite{Meessen,Hull,Hassan,Marolf1,Marolf2,Kachru} for 
the T-duality rules in presence of more general background.
\begin{eqnarray}
&& g_{xx} = {1\over j_{xx}} \cr
&& g_{\check i\check j} = j_{\check i\check j} \cr
&& e^{2\phi} = {e^{2\varphi}\over j_{xx}} \cr
&& H = {\cal H} \cr
&& F'_{n(x)} = F'_{(n-1)} \cr
&& F'_{n} = F'_{(n+1)(x)} \cr
&& {\gamma}^x = {\gamma}_x \cr
&& {\gamma}^{\check i} = {\gamma}^{\check i} ~.
\end{eqnarray}
Here we follow the notations of Ref.\cite{Kachru}. In particular,
$F'_n$ are gauge invariant RR field strengths and also $F_{n(x)}$ denotes
an $(n-1)$ form whose components are given
by
\begin{eqnarray}
\left[F_{n(x)}\right]_{i_1\cdots i_{n-1}} 
= \left[F_n\right]_{xi_1\cdots i_{n-1}} ~.
\end{eqnarray}

\noindent
{\large\bf A.6 ~~ The Mass Matrix}

In this section we will evaluate the fermion bilinear term due to the three 
form flux when the three brane wraps a divisor of $T^6$. Here we will use 
the coordinates $\{x^j,y^j\},~j= 1,\cdots,3$ to parametrize the spatial 
directions of the torus and $\{\hat x^j,\hat y^j\}$ for the corresponding 
tangent space indices. Now consider the relevant part of the action 
as given in Eq.(\ref{bilin}):
\begin{eqnarray}
S(\Theta^2) &=& - \mu_3\int d^4\zeta e^{-\phi} \sqrt{{\rm det}~g}
\Theta^T\Gamma^0\left({1\over 48} \Gamma^{mnp} H_{mnp}
- {1\over 16} \Gamma^{\tilde ipq} H_{\tilde ipq}
\right)\Theta \cr
&+& i {\mu_3\over 32} \int d^4\zeta  \sqrt{{\rm det}~g} \epsilon^{\tilde i\tilde j\tilde k\tilde l}
\Theta^T\Gamma^0\left({1\over 36}{\Gamma_{\tilde i\tilde j\tilde k\tilde l}}^{pqr} F'_{pqr}
+{\Gamma_{\tilde i\tilde j}}^p F'_{\tilde k\tilde lp}
%-{1\over 3} \bar\Theta{\Gamma_i}^{pq}\Theta F'_{jklpq}
\right)\Theta ~.
\end{eqnarray}
Here note that the first term in the second line vanishes for the case when the 
flux is turned on only along the compact directions. As a result we get
\begin{equation}
S(\Theta^2)={\mu_3\over 16} \int d^4\zeta \sqrt{{\rm det}~g}
\Theta^T\Gamma^0 \left(e^{-\phi}\left\{\Gamma^{\tilde ipq} H_{\tilde ipq}
-{1\over 3}\Gamma^{mnp} H_{mnp}\right\}
+ {i\over 2}  \epsilon^{\tilde i\tilde j\tilde k\tilde l} {\Gamma_{\tilde i\tilde j}}^p F'_{\tilde k\tilde lp}
\right)\Theta ~.
\end{equation}
In the following  we  will first consider the case when the three brane wraps the
divisor $Z^3 = constant$ and concentrate ourselves to the choice of flux
as given by Eq.(\ref{flux}). The above action  can be rewritten 
as 
\begin{eqnarray}
S(\Theta^2)={\mu_3\over 16} \int d^4\zeta \sqrt{{\rm det}~g}
~ \Theta^T\Gamma^0 M \Theta  ~,
\end{eqnarray}
with  the matrix $M$ defined to be
\begin{eqnarray}
\label{defM}
M = \left( e^{-\phi}\Gamma^{\tilde i\tilde jp} H_{\tilde i\tilde jp}
+ {i\over 2}  \epsilon^{\tilde i\tilde j\tilde k\tilde l} {\Gamma_{\tilde i\tilde j}}^p F'_{\tilde k\tilde lp}\right) ~.
\end{eqnarray}
The index $p$ now take value only along directions orthogonal to the divisor.

It is convenient to choose a basis, where the components of $\Theta$ are 
labelled as $|\epsilon_1,\epsilon_2,\epsilon_3,a>$, where $\epsilon_j = \pm 1,
~ j = 1,\cdots, 3~$ refer to ($-i$ times)
the eigen values of $\Gamma^{\hat x^j\hat y^j}$ 
respectively{\footnote{Here and in the following, the repeated index $j$ in
$\Gamma^{\hat x^j\hat y^j}$ as well as in
 $\left(\Gamma^{x^j}\Gamma^{y^j}\right)$ does not indicate 
a summation over $j$.}}. 
The label $a = \pm 1$ refers to the $SO(4)$ subgroup of the 
rotation group $SO(10)$. Before proceeding,  let us note here that from the commutation relations for the $\Gamma$ matrices it 
follows that $\Gamma^{\hat y^1\hat y^2\hat y^3}$ squares to $-1$ and as a result, 
$\Gamma^{\hat y^1\hat y^2\hat y^3} |\epsilon_1,\epsilon_2,\epsilon_3,a>$ can never vanish.

We will now evaluate the matrix, $M$, Eq.(\ref{defM}), in this basis. 
We start with the first term, $e^{-\phi}\Gamma^{ijp} H_{ijp}$ which arises from the DBI term.
{}From Eq.(\ref{flux}) it is easy to see that it takes the form, 
\begin{eqnarray}
\label{hmatrix}
M_{DBI}&=&e^{-\phi} \left [\Gamma^{x^1x^2x^3}-2\Gamma^{y^1y^2y^3}
-(\Gamma^{x^2x^3y^1} + 
\Gamma^{x^3x^1y^2}+\Gamma^{x^1x^2y^3}) \right.\cr &+& \left.
(\Gamma^{y^2y^3x^1}+\Gamma^{y^3y^1x^2}+\Gamma^{y^1y^2x^3}) \right] ~.
\end{eqnarray}
Here we note that the indices refer to the coordinate basis, which is different from the 
vierbein basis. 

The metric, with $r_a$ in Eq.(\ref{km}) set to unity takes the form
\beq
\label{metrica6}
ds^2=\sum_i|dx^i+\tau dy^i|^2
\eeq
where $\tau=e^{2\pi i\over 3}$. 
A convenient choice of vierbeins is then
\begin{eqnarray}
e_{x^1}^{\hat x^1} = 1, ~ e_{x^1}^{\hat y^1} = 0, ~
e_{y^1}^{\hat x^1} = \cos(2\pi/3), ~ e_{y^1}^{\hat y^1} = \sin(2\pi/3) ~.
\end{eqnarray}
The $\Gamma$ matrices in the vierbein basis and the coordinate basis are  related to each
 other by 
\begin{eqnarray}
&& \Gamma^{x^i} = \Gamma^{\hat x^i} - \cot(2\pi/3) \Gamma^{\hat y^i}~, ~
~ \Gamma^{y^i} = {\rm cosec}(2\pi/3) \Gamma^{\hat y^i}~, \cr
&&  \Gamma_{x^i} = \Gamma^{\hat x^i}~, ~ ~
\Gamma_{y^i} = \cos(2\pi/3) \Gamma^{\hat x^i} + \sin(2\pi/3) \Gamma^{\hat y^i}~.
\label{virco}
\end{eqnarray}
In particular one finds that 
$(\Gamma^{y^i})^2={\rm cosec}^2({2\pi\over 3}) ={4\over 3}$.
After some more algebra we can then write $M_{DBI}$ as,
\begin{eqnarray}
\label{Mdbitwo}
M_{DBI}&=&e^{-\phi} \left\{
%\left [
\left({3\over 4}\right)^3 \Gamma^{x^1}\Gamma^{y^1}
\Gamma^{x^2}\Gamma^{y^2}\Gamma^{x^3}\Gamma^{y^3} 
%-2 \right ] 
%\Gamma^{y^1}\Gamma^{y^2}\Gamma^{y^3}
%\cr &-& e^{-\phi} 
- \left({3\over 4}\right)^2\left[\Gamma^{x^2}\Gamma^{y^2}
\Gamma^{x^3}\Gamma^{y^3}
+\Gamma^{x^1}\Gamma^{y^1}\Gamma^{x^2}\Gamma^{y^2}
\right.\right. \cr &+& \left.\left.
\Gamma^{x^1}\Gamma^{y^1} \Gamma^{x^3}\Gamma^{y^3} \right] 
%\Gamma^{y^1}\Gamma^{y^2}\Gamma^{y^3}
%\cr &+& e^{-\phi}
+ {3\over 4}
\left[\Gamma^{x^1}\Gamma^{y^1} + \Gamma^{x^2}\Gamma^{y^2} 
+ \Gamma^{x^3}\Gamma^{y^3}\right]
-2\right\}\Gamma^{y^1}\Gamma^{y^2}\Gamma^{y^3} ~.
\end{eqnarray}

{}From Eq.(\ref{virco}) we get that 
\begin{eqnarray}
\left(\Gamma^{x^i}\Gamma^{y^i}\right) |\epsilon_1,\epsilon_2,\epsilon_3,a>
&=&{\rm cosec}\left({2\pi\over 3}\right) \left(\Gamma^{\hat x^i\hat y^i}
- \cot\left({2\pi\over 3}\right)\right) |\epsilon_1,\epsilon_2,\epsilon_3,a>
\cr &=&
{\rm cosec}^2\left({2 \pi\over 3}\right)
e^{{i\pi\over 3}\epsilon_i}|\epsilon_1,\epsilon_2,\epsilon_3,a> ~.
\label{eigen}
\end{eqnarray}

It then follows that $M_{DBI}$ acting on the state $|\epsilon_1,\epsilon_2,\epsilon_3,a>$ is
\beq
\label{mdbifi}
M_{DBI}|\epsilon_1,\epsilon_2,\epsilon_3,a>=\left({4\over 3}\right)   {\cal M}~ \Gamma^{\hat y^1\hat y^2\hat y^3} 
|\epsilon_1,\epsilon_2,\epsilon_3,a>
\eeq
where,  
\begin{eqnarray}
{\cal M} &=&
\left\{e^{- {i\pi\over 3}(\epsilon_1+\epsilon_2+\epsilon_3)} - 2
-e^{- {i\pi\over 3}(\epsilon_1+\epsilon_2)}
-e^{- {i\pi\over 3}(\epsilon_2+\epsilon_3)}
\right. \cr &-& \left.
e^{- {i\pi\over 3}(\epsilon_3+\epsilon_1)}
+ e^{-{i\pi\over 3}\epsilon_1}
+ e^{-{i\pi\over 3}\epsilon_2}
+ e^{-{i\pi\over 3}\epsilon_3}
\right\} ~.
\end{eqnarray}

Similarly we can evaluate the second term in Eq.(\ref{defM}) which arises due to the WZ terms, 
\beq
\label{defmwz}
M_{WZ} = {i\over 2} \epsilon^{\tilde i\tilde j\tilde k\tilde l} \Gamma_{\tilde i\tilde j}^p (F_{\tilde k\tilde lp} -C H_{\tilde k\tilde lp}) ~.
\eeq
Here $p$ takes values only over directions orthogonal to the divisor. 
It is easy to see that 
with an appropriate choice of orientation for 
the divisor \footnote{An opposite choice of orientation corresponds to a negative sign on the
 r.h.s below. This still gives the same number of zero modes.}, 
\begin{eqnarray}
{1\over 2} \epsilon^{\tilde i\tilde j\tilde k\tilde l} \Gamma_{\tilde i\tilde j}|\epsilon_1,\epsilon_2,\epsilon_3,a>
=  \epsilon_1\epsilon_2 \Gamma^{\tilde k\tilde l}|\epsilon_1,\epsilon_2,\epsilon_3,a> ~.
\end{eqnarray}
Thus, 
\beq
\label{actmwz}
M_{WZ}|\epsilon_1,\epsilon_2,\epsilon_3,a>=
i \epsilon_1\epsilon_2 \Gamma^{\tilde k\tilde lp} (F_{\tilde k\tilde lp} -C H_{\tilde k\tilde lp}) |\epsilon_1,\epsilon_2,\epsilon_3,a>.
\eeq
A little more algebra then shows that this can be written as,
\beq
\label{mwzifi}
M_{WZ}|\epsilon_1,\epsilon_2,\epsilon_3,a> ={4 i \over 3\sqrt{3}} \epsilon_1 \epsilon_2 \{
{\cal M}+1+e^{-{i\pi\over 3}(\epsilon_1+\epsilon_2+\epsilon_3)}\} \Gamma^{\hat y^1\hat y^2\hat y^3}
 |\epsilon_1,\epsilon_2,\epsilon_3,a>
\eeq
Adding, Eq.(\ref{mdbifi}), Eq.(\ref{mwzifi})  we finally  get that $M$ acting on 
$|\epsilon_1,\epsilon_2,\epsilon_3,a>$  
is given by, Eq.(\ref{actms}). 

As discussed above $ \Gamma^{\hat y^1\hat y^2\hat y^3} |\epsilon_1,\epsilon_2,\epsilon_3,a>$ cannot vanish. 
Thus the zero modes of $M$ can only arise if $m$, eq.(\ref{deflm}) vanishes. 
A quick inspection shows that  this happens when

a)    $\epsilon_1=\epsilon_2= \epsilon_3=1$  

b)  $\epsilon_1=\epsilon_2=\pm 1, \epsilon_3=-1$ 

c) $\epsilon_1=-\epsilon_2=\pm 1, \epsilon_3=-1$

Thus these values of $(\epsilon_1,\epsilon_2,\epsilon_3)$ give rise to zero modes.
Also $m$ does not vanish for any other choice of $(\epsilon_1,\epsilon_2,\epsilon_3)$.
Therefore there are no other zero modes.
Finally, one also finds that  the two states
$|\epsilon_1,\epsilon_2,\epsilon_3,a>$ and $|\epsilon_1,\epsilon_2,-\epsilon_3,a>$,
which are related by  $U(1)$ rotations in the $x^3-y^3$ plane, 
 have a different mass in general.
This is to be expected since the $U(1)$ symmetry is broken by the flux as discussed in 
section 5.  

Let us also now briefly discuss the case of the more general divisor $n_iZ^i=c$.
By relabelling the $Z^i$ coordinates if necessary we can always take $n^3\ne 0$. 
In this case it is useful to choose coordinates, $\psi^1,\psi^2,\psi^3$ which are related to 
the coordinates $Z^i$ as follows:
\begin{eqnarray}
\label{coordg}
&& Z^1=n_1 \psi^3 + n_3 \psi^1 \cr
&& Z^2=n_2 \psi^3 + n_3 \psi^2 \cr
&& Z^3=-n_1 \psi^1-n_2 \psi^2 + n_3 \psi^3~.
\end{eqnarray}  
$\psi^1,\psi^2$ are parallel to the divisor and $\psi^3$ is orthogonal to it. 
The divisor in these coordinates can be written as 
$\psi^3=constant$. 
The flux $G=F-\Phi H$ can be expressed as
\beq
\label{vg3}
G=const(d\psi^1\wedge d\psi^2\wedge d{\bar \psi}^3 + 
d\psi^2\wedge d\psi^3\wedge d{\bar \psi}^1 +
d\psi^3\wedge d\psi^1\wedge d{\bar \psi}^2)
\eeq
Upto a constant this is exactly the form of $G$ in the $Z^i$ coordinates. 
A further change of variables,
\begin{eqnarray}
&& {\tilde \psi}^1= \sqrt{n_3^2+n_1^2} \psi^1 + \sqrt{n_3^2+n_2^2} \psi^2 \cr
&& {\tilde \psi}^2=\sqrt{n_3^2+n_1^2} \psi^1 - \sqrt{n_3^2+n_2^2} \psi^2 \cr
&& {\tilde \psi}^3=\psi^3~,
\end{eqnarray}
preserves the form of $G$, Eq.(\ref{vg3}).
 It  also 
 allows the metric to be written in diagonal form as,
\beq
\label{metrgen}
ds^2=\sum_ir_i^2 |d{\tilde \psi}^i|^2.
\eeq
This is the same as the metric in the $Z^i$ coordinates we considered Eq.(\ref{km}). 
Thus the analysis for the general divisor 
maps after a change of coordinates to the case $Z^3=c$. 
And  we learn that for a general divisor also there are ten fermion zero modes.

\end{document}